\documentclass[acmsmall]{acmart}

\setcopyright{cc}
\setcctype{by-nc-nd}
\acmJournal{PACMHCI}
\acmYear{2026} \acmVolume{10} \acmNumber{2} \acmArticle{CSCW038}
\acmMonth{4} \acmPrice{} \acmDOI{10.1145/3788074}

\AtBeginDocument{%
  }

\usepackage{csquotes}
\usepackage{xspace}
\usepackage{svg}
\usepackage{tabularx}
\usepackage{graphicx}
\usepackage{pgf}
\usepackage{placeins}
\usepackage{tikz}
\usepackage{pifont}
\newcommand{\cmark}{\ding{51}}
\newcommand{\xmark}{\ding{55}}
\usepackage{soul}
\usepackage{array}

\usepackage[normalem]{ulem}

\usepackage{longtable}
\usepackage{graphicx}
\usepackage{booktabs}
\usepackage{caption}
\usepackage{setspace}
\usetikzlibrary{shapes.geometric}
\newcommand{\parabold}[1]{%
  \par\vspace{1ex}
  \noindent\textbf{#1. }%
}

\newcommand{\hlred}[1]{%
    {\sethlcolor{red!20}\hl{#1}}%
}
\newcommand{\hlgreen}[1]{%
    {\sethlcolor{green!20}\hl{#1}}%
}

\RenewDocumentEnvironment{quote}{+b}
  {\list{}{\leftmargin=1em\rightmargin=1em}\item\textit{\enquote{#1}}\endlist}
  {}

\newcommand{\toolname}{Shiksha Copilot\xspace}

\begin{document}

\title[Shiksha Copilot]{Shiksha Copilot: Teacher-AI Collaboration for Curating and Customizing Lesson Plans in Low-Resource Schools}

\author{Deepak Varuvel Dennison}
\email{dv292@cornell.edu}
\affiliation{%
  \institution{Cornell University}
  \city{Ithaca}
  \state{NY}
  \country{USA}
}

\author{Bakhtawar Ahtisham}
\affiliation{%
  \institution{Cornell University}
  \city{Ithaca}
  \state{NY}
  \country{USA}
}

\author{Kavyansh Chourasia}
\affiliation{%
  \institution{Microsoft Research}
  \city{Bengaluru}
  \country{India}
}

\author{Nirmit Arora}
\affiliation{%
  \institution{Microsoft Research}
  \city{Bengaluru}
  \country{India}
}

\author{Rahul Singh}
\affiliation{%
  \institution{Microsoft Research}
  \city{Bengaluru}
  \country{India}
}

\author{René F. Kizilcec}
\affiliation{%
  \institution{Cornell University}
  \city{Ithaca}
  \state{NY}
  \country{USA}
}

\author{Akshay Nambi}
\affiliation{%
  \institution{Microsoft Research}
  \city{Bengaluru}
  \country{India}
}

\author{Tanuja Ganu}
\authornote{Equal contribution.}
\affiliation{%
  \institution{Microsoft Research}
  \city{Bengaluru}
  \country{India}
}

\author{Aditya Vashistha}
\authornotemark[1]
\affiliation{%
  \institution{Cornell University}
  \city{Ithaca}
  \state{NY}
  \country{USA}
}

\renewcommand{\shortauthors}{Varuvel Dennison et al.}

\begin{abstract}
This study investigates \toolname, an AI-assisted lesson planning tool deployed in government schools across Karnataka, India. The system combined LLMs and human expertise through a structured process in which English and Kannada lesson plans were co-created by curators and AI; teachers then further customized these curated plans for their classrooms using their own expertise alongside AI support. Drawing on a large-scale mixed-methods study involving 1,043 teachers and 23 curators, we examine how educators collaborate with AI to generate context-sensitive lesson plans, assess the quality of AI-generated content, and analyze shifts in teaching practices within multilingual, low-resource environments. Our findings show that teachers used \toolname both to meet administrative documentation needs and to support their teaching. The tool eased bureaucratic workload, reduced lesson planning time, and lowered teaching-related stress, while promoting a shift toward activity-based pedagogy. However, systemic challenges such as staffing shortages and administrative demands constrained broader pedagogical change. We frame these findings through the lenses of teacher-AI collaboration and communities of practice to examine the effective integration of AI tools in teaching. Finally, we propose design directions for future teacher-centered EdTech, particularly in multilingual and Global South contexts.
\end{abstract}



\received{May 2025}
\received[revised]{November 2025}
\received[accepted]{December 2025}

\maketitle

\section{Introduction}
In many low-resource schools across the Global South, 
teachers often work in environments with overcrowded classrooms, staff shortages, and heavy administrative duties. 
These pressures contribute to time scarcity, excessive workloads, and teacher burnout~\cite{creagh_workload_2025, parihar2016study, kim_teachers_2019}. 
These challenges are compounded when teachers are required to teach multiple subjects across different grade levels, with limited institutional and pedagogical support.
In such contexts, developing lesson plans is a critical tool for structuring instruction and ensuring students meet learning goals.
Yet, the very constraints teachers face make it especially difficult to create high-quality lesson plans \cite{yuan_promoting_2016, dorji_teachers_2022}, despite their proven role in improving teaching effectiveness and student outcomes \cite{piper_effectiveness_2018}.
Beyond guiding instruction, lesson plans also serve a bureaucratic function, providing written evidence of preparation and curricular compliance required by school administrators and education authorities \cite{farrellLessonPlanning2002}. These structural constraints, combined with the critical role of lesson plans in both pedagogy and accountability, underscore the need for robust support systems to help teachers design and manage high-quality lessons. 

Large Language Models (LLMs) have shown great promise for generating high-quality lesson plans across subjects and languages~\cite{Rashid_Atilgan_etal._2024}. LLM-based tools like Khanmigo\footnote{https://www.khanmigo.ai/teachers} and MagicSchool AI\footnote{https://www.magicschool.ai/} have already made these capabilities available to teachers, providing them pedagogical support and ready-made templates to ease their workload. 
However, despite the promise and reach of these tools, significant knowledge gaps remain which are central to CSCW research. 
First, while LLMs produce fluent text, they often fall short on factual accuracy, cultural sensitivity, and contextual alignment \cite{Gallegos_Rossi_etal._2024, Huang_Yu_etal._2024}, 
particularly for languages and contexts in the Global South~\cite{Gallegos_Rossi_etal._2024, mirza_global-liar_2024}.  
Even when outputs are accurate, 
teachers must still interpret, adapt, and integrate these outputs into classroom practices \cite{Bitkina_Jeong_etal._2020, oecd2024ai}, a process that is shaped by institutional mandates and peer practices. 
Furthermore, we know little about how they perform in practice, particularly in Global South, where infrastructure, language, and pedagogical support differ sharply from Western contexts. 


This paper presents the design and development of an LLM-based lesson planning tool and examines how teachers in India used it in practice, focusing on collaboration, implementation challenges, and its influence on their work practices. We designed, deployed, and evaluated \toolname, a human-in-the-loop AI system to support teachers create curriculum-aligned, contextual lesson plans (LPs) in English and their local language. The tool was deployed in Karnataka, India, between December 2024 and March 2025,  to support grade 5–10 teachers to create LPs in English and Kannada. The system employs a structured workflow: LLMs generate initial content using Retrieval-Augmented Generation (RAG) frameworks; human curators then review these drafts for pedagogical quality, accuracy, and contextual relevance; teachers receive the curated materials and adapt them for classroom implementation. Teachers can also utilize supplementary AI features including curriculum-indexed chatbots for additional pedagogical support.

Drawing on mixed-methods data from 1043 teachers—including system logs, interviews, and surveys—we analyze how teachers interact with AI-generated materials, how they adapt them to local classroom needs, and how this process reshapes their work. In particular, we ask: 

\begin{itemize}
    \item[\textbf{RQ1:}] How do teachers collaborate with AI in an LLM-based lesson planning system?
    \item[\textbf{RQ2:}] What challenges and opportunities arise when implementing LLM-generated lesson plans in non-English medium classrooms?
    \item[\textbf{RQ3:}] In what ways does an LLM-based lesson planning tool reshape the work practices of the teachers?
\end{itemize}

Our findings show that teachers used AI-generated LPs for both classroom instruction and bureaucratic documentation. \toolname significantly reduced the time and effort spent on record-keeping while enabling access to comprehensive, curriculum-aligned plans. This led to measurable reductions in lesson planning time and teacher stress. We found high reliability in English LPs, while Kannada translations required extensive linguistic edits. However, in both cases, teachers actively drew on their own expertise and professional networks to address gaps, refine content, and ensure alignment with local curricular and classroom needs. Teachers also appropriated various features—such as question paper generators and AI chat—to support classroom practice. In doing so, the tool not only fit into existing workflows but also reshaped them, reflecting both the promise and limits of AI integration in low-resource settings. 

Building on our findings, we discuss what makes teacher–AI collaboration effective, emphasizing teachers’ active role as evaluators and adopters of AI-generated content. We highlight how AI-driven educational tools can be embedded within existing Communities of Practice to foster collective knowledge creation and ensure cultural and linguistic relevance. Finally, we reflect on the limitations of current LLMs for “low-resource” languages and argue for human-in-the-loop and community-driven approaches to improve quality, equity, and sustainability of AI in education. In sum, our paper makes three contributions to CSCW:
\begin{itemize}
    \item We present the design and implementation of \toolname, a human-in-the-loop AI system that supports bilingual lesson planning, offering 
    insights into building Human-AI systems for linguistically diverse educational contexts.
    \item We provide empirical evidence of teacher–AI collaboration, demonstrating how teachers evaluate, adapt, and contextualize AI-generated content, highlighting both the strengths and limits of current LLMs.
    \item We show how AI tools transform teacher workflows in Global South classrooms, influencing both instructional practices and administrative processes, and driving organizational change. 

\end{itemize}

\section{Related Work}
Our work builds on two strands of CSCW and HCI research: (1) studies of teacher-facing technologies in resource-constrained educational settings in the Global South, and 
(2) the emerging use of AI, especially LLMs, in teaching and pedagogy. 
Together, these bodies of work 
provide a foundation for understanding how emerging AI tools are integrated into everyday classroom practices and adapted by teachers working in situated, often highly constrained, environments.

\subsection{Technology for Teachers in the Global South}

Technology is often positioned as a pathway to educational equity, providing teachers with access to new instructional strategies and professional development opportunities \cite{friedrich2013how}. However, such narratives frequently reflect techno-optimistic assumptions that overlook the need to align tools with local pedagogical practices, infrastructural constraints, and institutional realities \cite{mustafa2024challenges, chand2020why, warschauer2010olpc}. The One Laptop Per Child initiative, for instance, highlights how technological interventions can falter when they fail to consider teachers’ needs and classroom contexts \cite{warschauer2010olpc, wexler2021whylearning}. Designing effective technological solutions, therefore, requires centering the experiences and constraints of teachers. 

Across India, sub-Saharan Africa, and Latin America, educators have shown a willingness to adopt digital tools and often adapt them creatively to local contexts, even in the face of infrastructural deficits, language barriers, and top-down implementation models that limit meaningful integration \cite{kaur2018teachers, nasreen2018perception, mustafa2024challenges, chand2020why, agyei2013analysis}. For example, during the COVID-19 pandemic, Indian teachers used WhatsApp-based helpline to receive pedagogical and emotional support through 
trained facilitators \cite{varanasi2024saharaline}. Meghshala, a teacher-support organization, provided teachers with structured, multimedia-driven LPs that teachers could adapt to their classroom needs 
\cite{reddy2022meghshala}. Similarly, Open Educational Resources (OERs) offered scalable ways to support instruction—particularly when paired with training or peer mentorship \cite{walker2023trialling}. In India, initiatives like TESS-India showed that localized, high-quality OERs can enhance teacher education and classroom practice \cite{perryman2013tessindia}. 
However, the success of these initiatives ultimately hinges on more than just content availability; it requires adequate infrastructure, teachers’ digital skills, and materials that are culturally and linguistically relevant to their classrooms \cite{perryman2013tessindia, perryman2013oercommunities}.

With the emergence of generative AI, scholars argue that technologies such as LLMs can further enhance the OER ecosystem by automating content creation and enabling personalization \cite{bozkurt2023generative}. Some suggest that with GenAI’s ability to generate customized LPs, assessments, and explanations, the need for static, pre-curated OERs may diminish \cite{aksoy2024behind}. However, this raises concerns about the reliability, pedagogical soundness, and ethical use of AI-generated materials—especially in low-resource settings where educators may lack the time or expertise to evaluate them \cite{aksoy2024behind, tlili2022unleashing}. Rather than replacing OERs, a promising direction involves leveraging GenAI’s flexibility to adapt and extend high-quality, openly licensed resources, while ensuring human oversight, quality control, and alignment with local pedagogical contexts \cite{bozkurt2024genai, tlili2022unleashing}.

Beyond tool access, teacher professional development remains a persistent challenge in the Global South. Although many teachers work in communities similar to their own, training programs often overlook the sociocultural and infrastructural complexities of classroom environments, including multilingual instruction, resource constraints, and large class sizes \cite{anwaruddin2016ict, leach2005deepimpact, marcondes1999brazil, raina1999indigenizing, villegas2003teacherdev, tsp2020}. Centralized curricula further restrict pedagogical autonomy, limiting teachers’ ability to adapt content to local needs \cite{tsp2020, samantha2013}. 
Prior work has emphasized that culturally responsive technology integration requires grounding design and training in teachers’ lived experiences, rather than assuming uniform uptake \cite{ogan2019grain, uchidiuno2019learning}. Our work builds on these insights by examining how teachers in Indian schools collaborate with an AI-based tool to create LPs. While prior interventions have explored how teachers adapt pre-authored content, the emergence of generative AI shifts the focus toward co-creating instructional materials in real time. In the following section, we examine recent work on AI tools for supporting teachers, with an emphasis on lesson planning, and the evolving dynamics of teacher–AI collaboration.

\subsection{AI For Supporting Teachers}

AI is increasingly being adopted in educational settings to assist teachers with a range of tasks, including grading, administrative support, generating instructional content, and providing personalized feedback \cite{wang2024survey,oecd2024ai,ai25-1162}. These applications promise to reduce teacher workload, improve instructional quality, and support differentiated instruction. However, their effectiveness depends heavily on curricular alignment, contextual integration, and teacher trust \cite{viberg2024trust,bitkina2020trust}. At the same time, \citet{woodruff2024knowledge} caution that introducing AI into knowledge work can reconfigure labor and deskill practitioners. This is particularly relevant for teachers who make extensive use of AI in their knowledge work. 

Early tools such as intelligent tutoring systems and automated rubric-based assessments laid the foundation for AI-assisted teaching \cite{anagnostopoulou2024trustworthy}. More recently,  LLMs, such as ChatGPT, have emerged as generative tools capable of producing quizzes, comprehension exercises, and even complete LPs \cite{xiao2023reading,baytak2024content,rashid2024readability}. Studies show that LLM-generated LPs often resemble those created by experienced educators and can be particularly helpful for novice teachers, offering pedagogical support and reducing preparation time \cite{vandenberg2023chatgpt,karaman2024chatgptplans}. 

Yet these benefits are accompanied by well-documented limitations. Research has shown that LLMs can hallucinate, generate fabricated citations, and struggle to adapt content to local pedagogical norms \cite{sridhar2023curricular,powell2024firstgrade,dornburg2024chatgpt}. \citet{harveyDontForgetTeachers2025}, for example, examine educator-centered perspectives on LLM-related harms and their broader implications. 
Studies of human–AI collaboration in educational contexts report mixed or limited impact \cite{kuechemann2025lmfms}, highlighting the need for more context-sensitive design that genuinely supports teachers’ practices and environments \cite{molenaar2022hybrid}. 
Comparative evaluations also show that AI-generated content often reinforces rigid teaching formats and rarely accommodates learner diversity or cultural specificity \cite{durmus2024realclass,clark2024overlords}. 

In response to these challenges, researchers have developed structured co-design frameworks to better align AI outputs with classroom needs. 
For instance, LessonPlanner embeds Gagné’s instructional principles into interactive UIs \cite{fan2024lessonplanner}, while RAG frameworks refine LLM outputs through iterative feedback loops \cite{tan2025elf,zheng2024automatic,mullins2024rag}. Fine-tuned models and mega-prompting strategies that incorporate learner profiles, instructional goals, and teaching styles have also improved output fidelity \cite{karpouzis2024genai,koenig2024finetuning}. However, despite these advances, most research and design efforts to use AI to improve teachers' skills and capacity remain primarily concentrated in Western, English-dominant contexts \cite{bucchiarone2024lessonplans}. In multilingual and low-resource classrooms, these systems often struggle because of lack of localized datasets, insufficient language support, and unique challenges faced by the teachers \cite{sebastian2024emerging}. 

Research also emphasizes that effective AI integration requires not only improving AI content quality, but also fostering productive Teacher-AI Collaboration (TAC) \cite{sun2024teacherai}. TAC is often framed through two perspectives: a replacement model where AI substitutes the teacher in delivering instruction or assessments, and an augmentation model that supports teachers with content creation, feedback, and personalization \cite{jeon2023llm,baker2016stupid}. Building on this augmentation view, recent theory argues that how we frame AI---as a tool versus a collaborator---shapes whose agency is amplified. \citet{satyanarayan2024agency} propose treating humans and GenAI as co-agents whose compositional and interpretive roles can be dynamically reconfigured, with “design as delegation of constrained agency” and “ethics as care” guiding longer-term stewardship. For example, \toolname, the teacher-AI collaborative system that we study, follows this co-agency stance, keeping teachers accountable decision-makers while letting the model assume bounded, negotiable roles in drafting and refinement. 


\textbf{Empirical studies in the Global South} further illustrate how TAC depends on broader ecosystems involving designers, content curators, and institutional intermediaries. For example, Choi et al. \cite{choi2024llms} found that while an AI chatbot improved teacher confidence in Sierra Leone, infrastructural limitations and trust gaps hindered sustained use. In India, educators valued the time-saving potential of generative tools but faced usability barriers including long response times, absence of vernacular content, and misalignment with school routines \cite{sebastian2024emerging,santos2024llmindia}.  Our work extends this emerging literature by examining teacher-AI collaboration in Indian classrooms, where educators must adapt LLM outputs to multilingual contexts, assess pedagogical fit, and mediate their role as instructional decision-makers. In addition to exploring the questions of LLM-system accuracy, we explore how teachers embed AI into their everyday workflows and discuss 
the social, cultural, organizational, and infrastructural dimensions of teacher–AI interaction.

\section{System Design and Implementation}

\toolname was developed to support teachers in low-resource Indian classrooms by enhancing lesson planning workflows through an LLM-based system with human-in-the-loop curation and bilingual support. The system was developed in close collaboration with an implementation partner that runs several initiatives to strengthen public education in Karnataka, India for over 22 years. \toolname was piloted in Karnataka as part of the organization’s larger vision of making quality teaching and learning resources accessible to English- and Kannada-medium public school teachers. Aside from \toolname, the interventions designed by the partner organization are implemented on the ground by mentors, who serve as conduits between the organization’s program team and public school teachers. These mentors train the teachers and provide ongoing support across all programs. The same approach was followed for \toolname. The costs for lesson plan curation, tool deployment with teachers, and LLM API usage were funded by the partner organization through philanthropic grants.


To inform the design, the partner conducted focus-group discussions with 29 teachers in batches to identify key pain points and opportunities where AI-based tools could offer meaningful support. The partner organization surfaced three critical needs through these consultations: alignment with the state-mandated curriculum, scaffolding for low-resource classrooms with limited digital infrastructure and learning materials, compliance with the 5E Instructional Model.

The 5E Instructional Model is a constructivist framework for teaching that guides learners through five phases: \emph{Engage}, \emph{Explore}, \emph{Explain}, \emph{Elaborate}, and \emph{Evaluate}. In the \emph{Engage} phase, teachers spark curiosity and connect prior knowledge to new ideas; in \emph{Explore}, students investigate phenomena through inquiry and hands-on activities; in \emph{Explain}, they articulate understanding while teachers formalize concepts; during \emph{Elaborate}, students apply knowledge to new contexts; and in \emph{Evaluate}, both teachers and students assess learning and reflect on progress \cite{polanin_effects_2024}.

The 5E format is mandated by the Karnataka education department, and contextual adaptability for diverse teaching environments. In response to these inputs, the system was designed to prioritize curriculum-aligned content generation and support contextualized teaching across both English and Kannada mediums. Its design was grounded in the following principles:
\begin{itemize}
\item \textbf{Curriculum Alignment:} Ensure all LPs closely follow the state-mandated syllabus and grade-level learning objectives. 
\item \textbf{Bilingual Support:} Provide localized content for both English- and Kannada-medium classrooms.
\item \textbf{Low-Tech Accessibility:} Enable mobile and offline access to accommodate infrastructure constraints.
\item \textbf{Workflow Integration:} Align outputs with administrative documentation, pedagogy standards (e.g., 5E model), and classroom realities.
\item \textbf{Human-in-the-Loop Curation:} Engage educators as reviewers and co-designers to ensure pedagogical rigor and contextual relevance.
\end{itemize}

\subsection{System Design}
The technical architecture of \toolname is grounded in the core design principles outlined earlier and
comprises three key components: \textbf{Curriculum Ingestion}, \textbf{Translation}, and \textbf{LLM Orchestration}.



\begin{figure}[t]
    \centering
    \includegraphics[width=0.9\linewidth]{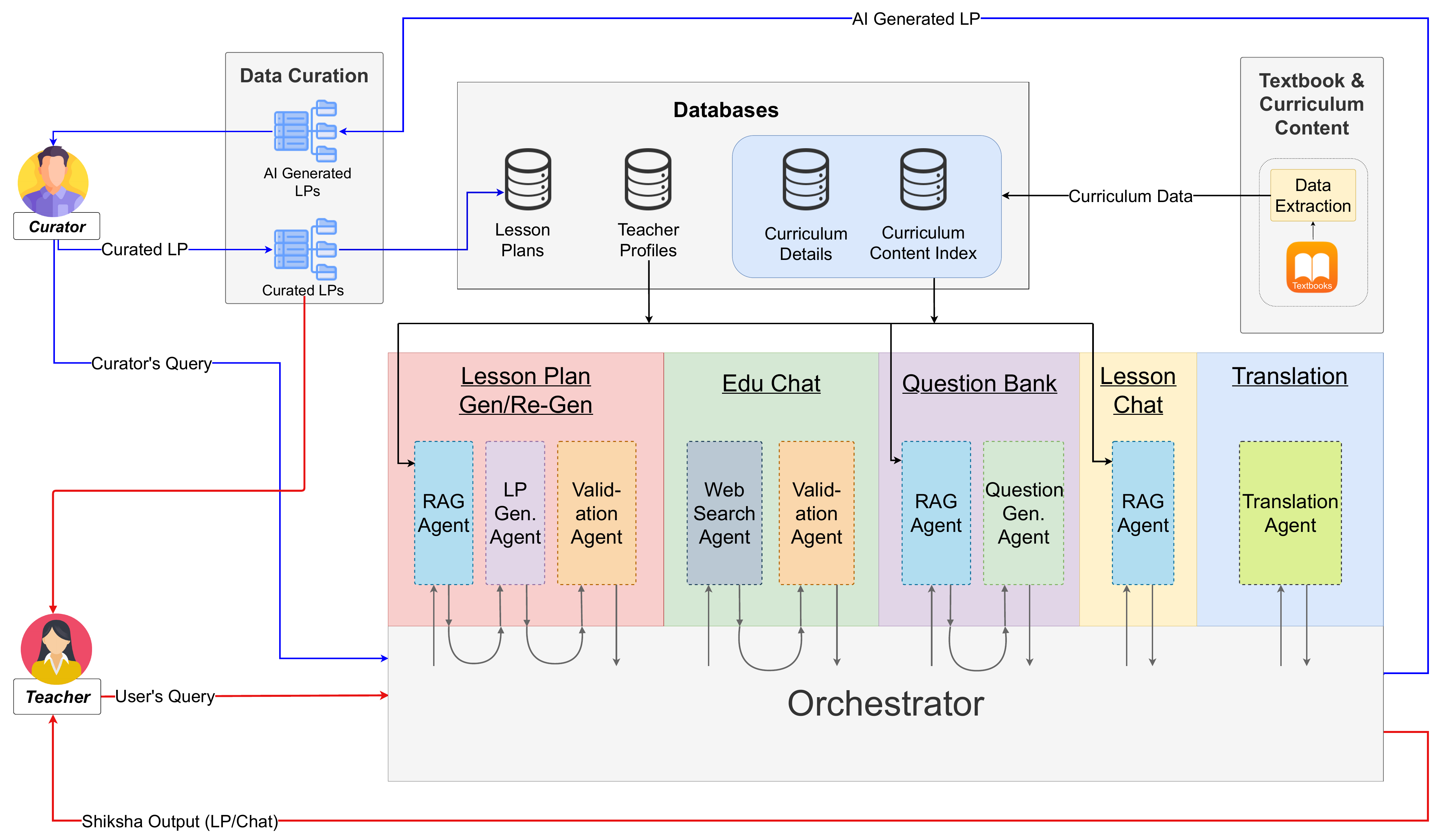}
    \Description[System flow diagram of \toolname]{Detailed system architecture diagram of the \toolname platform. On the left, teachers submit user queries and curators submit curation queries. Curriculum content is extracted from textbooks and stored as curriculum details and a curriculum content index in databases, alongside lesson plans and teacher profiles. Curators review and curate AI-generated lesson plans, which are stored back into the system. An orchestrator manages multiple AI agents, including retrieval-augmented generation agents, lesson plan generation and validation agents, web search and validation agents for Edu Chat, question generation agents for a question bank, and translation agents. These components work together to produce final \toolname outputs such as lesson plans and lesson chat responses delivered back to teachers.}
    \caption{System Architecture for \toolname}
    \label{fig:SystemArchitecture}
\end{figure}

\subsubsection{Curriculum Ingestion.}
Textbooks for four subjects (Science, Mathematics, Social Science, and English) across Grades 5–10 were ingested into the system. This involved extracting structured text and metadata from PDFs (e.g. chapters, subtopics, learning outcomes), and chunking and indexing chapter content. The resulting curriculum RAG index formed the foundation for textbook-grounded content generation.

Each textbook page was converted to images using pdf2image\cite{pdf2image}, from which the Table of Contents metadata was extracted via \emph{OpenAI GPT-4o}\cite{openai_gpt4o_system_card} LLM model. Next, optical character recognition was performed using \emph{Tesseract}\cite{tesseract_ocr} and \emph{Azure Form Recognizer}\cite{azure-form-recognizer}, with chapter-wise segmentation.
Extracted text is chunked (512 tokens per chunk) and embedded using OpenAI’s \emph{ada-embedding}\cite{openai_ada} model. These embeddings were stored in \emph{Azure AI Search}\cite{azure_ai_search}, indexed by chapter, enabling efficient retrieval during downstream inference.

\subsubsection{Translation.}
To support bilingual content, we included a dedicated translation component for English-Kannada translation with a fine-tuned translation model optimized for education domain. The translation model is fine-tuned using Low-Rank Adaptation (LoRA)\cite{hu2021lora} on top of \emph{Sarvam-1}\cite{sarvam1}, a decoder-only LLM pre-trained for Indic languages. To address the specific translation needs in \toolname like translation of long multi-paragraph content, understanding of domain specific educational terms and lesson plan and chat formatting, following types of data was used for fine-tuning:

\begin{itemize}
    \item Textbook-aligned English-Kannada sentence pairs from English and Kannada textbooks (using GPT-4-assisted bilingual alignment)
    \item Synthetic sentence generation using LLM based on educational glossary 
    \item Human-annotated general examples for English-Kannada pairs from IndicTrans2\cite{gala2023indictrans}
    \item Human-annotated LPs examples, put together by the local teachers working with the partner organization.
\end{itemize}





\subsubsection{LLM Orchestration.}
The LLM Orchestrator was designed as a central engine to orchestrate various AI-driven features in \toolname shown in Table \ref{tab:features-by-role}. It was designed as an agentic framework driven by (\emph{OpenAI 4o model}\cite{openai_gpt4o_system_card}) LLM, with Retrieval-Augmented Generation (RAG)\cite{lewis2020rag} as the backbone. 
The orchestrator includes a set of reusable agents, each designed to fulfill a specific instructional role. 

\begin{itemize}
    \item The \emph{RAG agent} retrieves relevant content from the curriculum index to contextualize LPs, chatbot responses, and assessments.
    \item The \emph{Lesson Plan Generation agent} sequentially constructs LPs aligned to the 5E model (Engage, Explore, Explain, Elaborate, Evaluate), ensuring continuity and thematic coherence.
    \item The \emph{Question Generation agent} automates the creation of assessment items using Bloom’s Taxonomy, allowing teachers to specify chapter coverage and cognitive depth.
    \item The \emph{Lesson Chat} and \emph{Edu Chat agents} support teacher queries, with the former grounded in lesson-specific content and the latter offering general academic help.
    \item To ensure bilingual accessibility, the \emph{Translation agent} converts English LPs into Kannada, supporting consistent formatting and pedagogical clarity based on the fine-tuned SLM translation model mentioned earlier. 
    \item A \emph{Validation agent} monitors formatting, completeness, and model responses, enabling controlled regeneration and fallback strategies.
    
\end{itemize}







Together, these agents form a modular and extensible orchestration layer that powers all user-facing features, from AI-assisted editing to curriculum-aligned Q\&A. This agentic approach promotes reusability, interpretability, and maintainability—critical for iterative deployment in low-resource, multilingual educational settings.


\subsection{System Implementation}
\begin{figure}[t]
  \centering
    \includegraphics[width=1.0\linewidth]{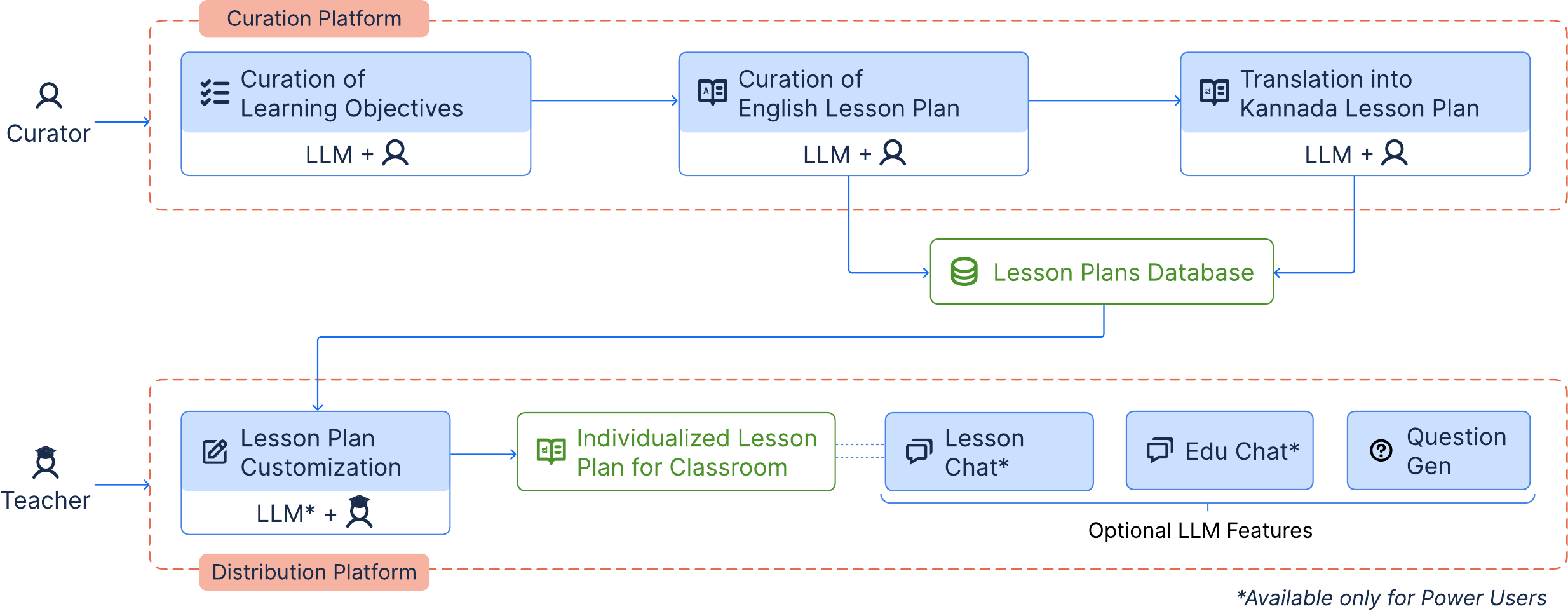}
    \Description[System workflow diagram]{System workflow diagram divided into a curation platform and a distribution platform. In the curation platform, a curator works with an LLM to curate learning objectives, develop an English lesson plan, and translate it into a Kannada lesson plan. These lesson plans are stored in a central lesson plans database. In the distribution platform, teachers use an LLM to customize lesson plans, resulting in individualized lesson plans for the classroom. Optional LLM-based features include lesson chat, educational chat, and question generation, which are available only to power users.}
  \caption{System Workflow of \toolname}
  \scriptsize{Learning Objectives are first generated using LLMs and reviewed by curators. These are used to generate English lesson plans (LPs), which undergo further review before being translated into Kannada and reviewed again by curators. The curated LPs are stored in a shared database. Teachers access these plans through the distribution platform and create their own copies for classroom use. Each customized lesson plan is also linked to a Lesson Chat, enabling teachers to ask lesson-specific queries.}
  \label{fig:SystemWorkflow}
\end{figure}

The \toolname system was implemented in two phases: a \textbf{Curation Phase} and a \textbf{Distribution Phase}. In the Curation Phase, curators—teachers with specific subject expertise—worked with the AI system to develop high-quality, error-free base LPs. In the Distribution Phase, these curated LPs were made available to teachers, who could customize them for their own classrooms. This two-phase process balanced quality control with classroom adaptability. Each phase is described in detail in the sections that follow. Figure~\ref{fig:SystemWorkflow} illustrates the overall system workflow. 

\subsubsection{Curation Phase.}
\begin{figure}[ht]
    \centering
    \includegraphics[width=\linewidth]{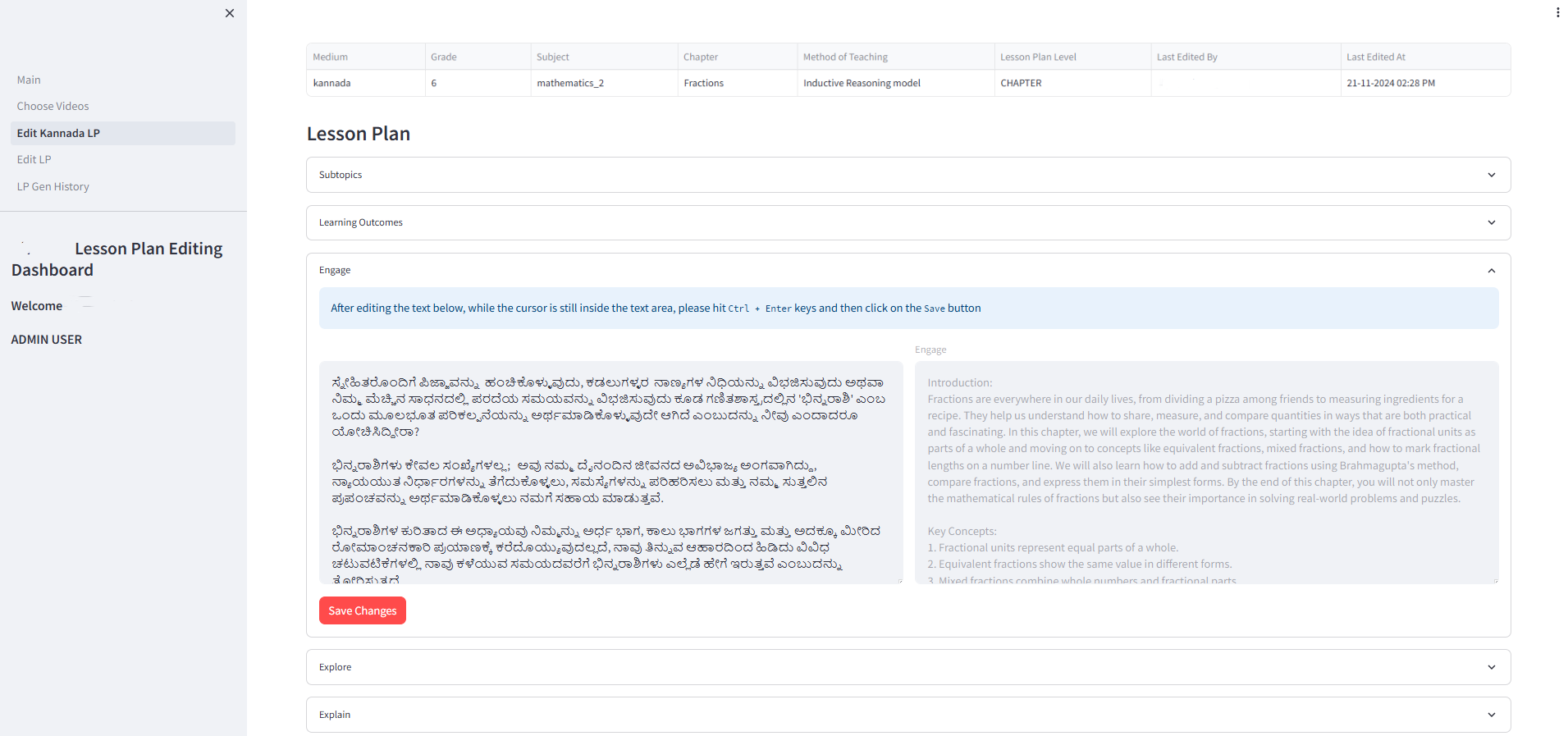}
    \Description[User interface of the curator platform]{User interface screenshot of a lesson plan editing dashboard for a Kannada-medium Grade 6 mathematics lesson on fractions. At the top, metadata fields display medium, grade, subject, chapter, teaching method, lesson plan level, and last edited details. The main content area shows expandable lesson plan sections including subtopics, learning outcomes, and engage. The engage section is open and includes a Kannada text editing area on the left and an English reference version on the right. Instructions for saving edits and a prominent save changes button are visible. A navigation sidebar on the left provides access to lesson plan editing, history, and admin options.}
    \caption{Curator's Platform for Editing LPs}
    \label{fig:curatorLPView}
\end{figure}

In the Curation Phase, a team of 23 curators—15 English-medium and 8 Kannada-medium educators—used the web-based Curation Platform (as shown in Figure \ref{fig:curatorLPView}) in \toolname to develop LPs. Curators were selected by the pedagogy team of the partner organization. They included a mix of experienced teachers and student educators with backgrounds in working with public school students. All curators held at least a bachelor’s degree in the relevant subject area for which they developed lesson plans. The process began with English-medium LPs: LLMs first generated draft Learning Objectives (LOs), which were reviewed and augmented by English-medium curators to ensure comprehensive curricular coverage and alignment with prescribed standards. These curated LOs then seeded LP generation via LLMs enhanced with ingested curriculum content through RAG. To account for the material constraints of target classrooms, prompts were designed to generate content suitable for low-resource settings.

AI-generated LPs were organized into discrete content blocks—covering the five phases of the 5E framework, suggested activities, assessment questions, and real-world applications. Curators reviewed each block to ensure accuracy, pedagogical clarity, and classroom feasibility. Blocks were individually rated, revised, and assembled into finalized LPs.

These finalized English LPs were translated into Kannada using the translation module. Kannada-medium curators then reviewed the translations, focusing on grammar, terminology, and stylistic consistency. Once curated in both languages, the LPs were made available to teachers for classroom use. A sample curated plan is included in the Appendix~\ref{appendix:sample-lp}.

\begin{table}[!ht]
  \centering
  \small
  \caption{Feature availability by user role}
  \label{tab:features-by-role}
  \begin{tabular}{|
      >{\raggedright\arraybackslash}p{5cm}|
      >{\centering\arraybackslash}p{1.5cm}|
      >{\centering\arraybackslash}p{1.5cm}|
      >{\centering\arraybackslash}p{1.5cm}|
    }
    \hline
    \textbf{Features}
      & \textbf{Curators}
      & \textbf{Power Users}
      & \textbf{Standard Users} \\
    \hline

    Review \& refine AI-generated LOs
      & \cmark & \cmark & \cmark \\
    \hline

    Review \& refine AI-generated LPs
      & \cmark & \cmark & \cmark \\
    \hline

    Manual LP Modification
      & \cmark & \cmark & \cmark \\
    \hline

    AI-assisted LP Modification
      & \xmark & \cmark & \xmark \\
    \hline

    AI-based Regeneration of LPs on LO change or with teacher feedback
      & \cmark & \cmark & \xmark \\
    \hline

    Lesson Chat (Q\&A for Lesson-specific support)
      & \xmark & \cmark & \xmark \\
    \hline

    Edu Chat (General Q\&A for academic support)
      & \xmark & \cmark & \xmark \\
    \hline

    Question Gen (AI-based Question Paper Generation)
      & \xmark & \cmark & \cmark \\
    \hline
  \end{tabular}
\end{table}

\subsubsection{Distribution Phase.}\label{subsub:distribution-phase}
\begin{figure}[H]
    \centering
    \includegraphics[width=\linewidth]{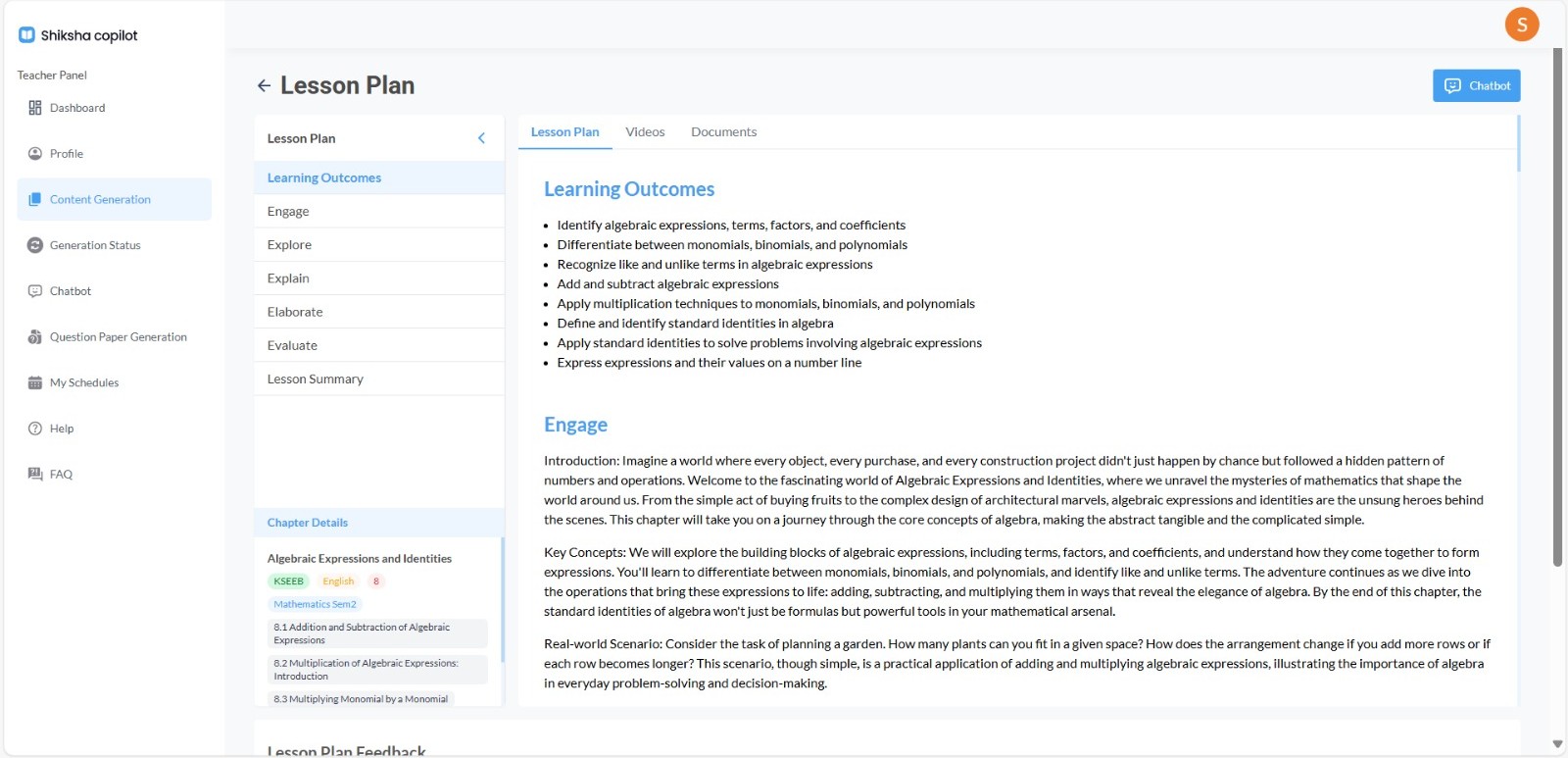}
    \Description[User interface screenshot of the teacher's platform]{User interface screenshot of a teacher lesson plan view. A navigation sidebar on the left includes options such as dashboard, profile, content generation, chatbot, and question paper generation, along with lesson sections like learning outcomes, engage, explore, explain, elaborate, evaluate, and lesson summary. The main panel displays a lesson plan for algebraic expressions and identities, showing detailed learning outcomes in bullet form and an engage section with an introduction, key concepts, and a real-world gardening scenario. Chapter details and subtopics are visible in a side panel, and a chatbot button appears in the top-right corner.}
    \caption{LP View in the Teacher's Platform}
    \label{fig:teacherLPView}
\end{figure}

In the Distribution Phase, \toolname was deployed with 1,043 field teachers working in Kannada- and English-medium public schools across Karnataka, India. All the schools were Kannada-medium, and a subset of them also included English-medium classrooms. Most schools were low-resource, with limited digital infrastructure and a shortage of high-quality teaching and learning materials. The partner organization collaborated with government officials to recruit teachers from 757 schools across 35 educational districts, ensuring statewide coverage. The officials played a critical role in identifying teachers for training and pilot implementation. Teachers were informed about the study prior to training and provided consent to participate through a pre-survey. They were also informed of their right to withdraw at any time, with clear instructions for doing so included in the consent form. Only the teachers who participated in the study had access to the tool. Teachers accessed curated LPs via a web-based Distribution  (as shown in Figure \ref{fig:teacherLPView}, where they could create and customize their own copies.

Deployment was phased, with teachers receiving a three-hour training led by four mentors from the partner organization. Each mentor supported approximately 250 teachers and conducted around ten training sessions, held either in person or virtually. Mentors also monitored usage and offered ongoing support via a dedicated WhatsApp group, phone calls, and in-person visits.

\parabold{User Roles and Feature Access}
Initially, all teachers were given \textbf{standard user} access. Standard users could view and manually customize curated LPs and use the AI-assisted \emph{Question Gen} feature, which generated assessments aligned with curricular blueprints based on Bloom’s Taxonomy \cite{forehand2010bloom}. Based on platform engagement, 146 teachers were granted \textbf{power user} access by the partner organization. This access was rolled out sequentially to gather structured feedback on the AI features from teachers and to build greater enthusiasm, drawing on the organization’s prior experience of working with them. In addition, the cost of AI usage was taken into account to ensure that the organization could plan for and manage expenses effectively based on actual usage patterns.  In addition to the standard features, power users could generate entirely new LPs using AI and apply AI-driven edits to curated LPs. They also had access to two interactive chatbot tools. \emph{Lesson Chat} was automatically linked to each lesson plan, allowing teachers to ask lesson-specific questions such as concept clarifications or activity suggestions. \emph{Edu Chat}, on the other hand, served as a general-purpose educational assistant for broader pedagogical support, including strategies for differentiated instruction.

Table \ref{tab:features-by-role} provides details on the availability of key system features by user role. Curators had full rights for creating and translating LPs. Power users had access to all AI tools for lesson editing, chat, and quiz generation, whereas standard users could view materials, manually customize them, and use the question paper generation.

\section{Methods}
To understand how teachers collaborate with \toolname to generate LPs and how \toolname reshapes their everyday work practices, we used various data collection methods such as semi-structured interviews, surveys, and system logs. The study protocol was Institutional Review Board (IRB) approved, and all procedures adhered to ethical research guidelines.
\begin{figure}[t]
    \centering
    {\begin{tikzpicture}[x=\textwidth/11.5, font=\sffamily] 

\newcommand{\schoolevent}[3]{%
  \node[diamond, draw=red!50!black, fill=red!40, inner sep=1.5pt, minimum size=1pt] at (#1,#2) {};
  \node[above right, align=center, rotate=90, yshift=-6.5pt, xshift=2px] at (#1,#2) {\scriptsize #3};
}

\newcommand{\studyevent}[3]{%
  \node[star, draw=orange!50!black, fill=orange!40, inner sep=1.5pt, minimum size=1pt] at (#1,#2) {};
  \node[above right, align=center, rotate=90, yshift=-6.5pt, xshift=2px] at (#1,#2) {\scriptsize #3};
}

    \node[diamond, draw=red!50!black, fill=red!40, inner sep=1.5pt, minimum size=1pt] at (1,-1.3) {};
    \node[above right, align=center] at (1,-1.5) {\footnotesize School Timeline};

    \node[star, draw=orange!50!black, fill=orange!40, inner sep=1.5pt, minimum size=1pt] at (3.5,-1.3) {};
    \node[above right, align=center] at (3.5,-1.55) {\footnotesize Study Milestones};

    \node[rectangle, draw=blue!40!black, fill=blue!30, inner sep=1.5pt, minimum size=4pt] 
    at (5.75,-1.3) {};
    \node[above right, align=center] at (5.75,-1.5) {\footnotesize Lesson Plan Curation};

    \node[rectangle, draw=green!40!black, fill=green!30, inner sep=1.5pt, minimum size=4pt] 
    at (8.45,-1.3) {};
    \node[above right, align=center] at (8.45,-1.55) {\footnotesize Tool Deployment};

    \fill[blue!20, fill opacity=0.8] (1, -0.2) rectangle (4.5, 0.2);

    \fill[green!20, fill opacity=0.8] (5, -0.2) rectangle (9, 0.2);

  \draw[thick] (0,0) -- (11,0);
  
  \foreach \i/\m/\y in {
    0/Jul/2024,
    1/Aug/2024,
    2/Sep/2024,
    3/Oct/2024,
    4/Nov/2024,
    5/Dec/2024,
    6/Jan/2025,
    7/Feb/2025,
    8/Mar/2025,
    9/Apr/2025,
    10/May/2025,
    11/Jun/2025
  }{
    \draw[thin, gray] (\i,3pt) -- (\i,-3pt);
    \node[below=6pt, align=center] at (\i,0) {\scriptsize \m \\[-3pt]\scriptsize \y};
  }

\schoolevent{0.01}{0}{First Term Begins}
\schoolevent{5.0}{0}{Second Term Begins}
\schoolevent{9.0}{0}{Exam Season Begins}
\schoolevent{10.0}{0}{Summer Break Begins}

\studyevent{1.0}{0}{LP Curation Begins}
\studyevent{2.8}{0}{Curator Interviews}
\studyevent{4.3}{0}{Pre-survey Distribution}
\studyevent{6.65}{0}{Teacher Interviews}
\studyevent{8.25}{0}{Post-survey Distribution}
\studyevent{8.5}{0}{Log Analysis Data Cutoff}

\end{tikzpicture}}
    \label{fig:StudyTimeline}
    \caption{School \& Study Timeline}
\end{figure}
\subsection{Semi-Structured Interviews}
We conducted semi-structured interviews with three key stakeholder groups: (1) curators who reviewed and refined AI-generated LPs, (2) teachers who used the Distribution Platform for their lesson planning, and (3) mentors who supported teachers during the deployment. The interview protocols were tailored to each group based on their roles and were designed to address our core research questions: how teachers engage with AI (RQ1), the usability and effectiveness of LLM-generated LPs (RQ2), and the broader impact on teachers' work practices (RQ3).

\parabold{Recruitment and Participants}
A total of 43 interviews were conducted: 17 expert curators (13 English-medium, 4 Kannada-medium), 22 teachers (10 Power users and 12 Standard users), and all four mentors who worked closely with teachers during the deployment period. Curators were selected to ensure representation across subjects and grade levels in both English and Kannada. Teachers were recruited to capture diversity across subject domains including English, Science, Mathematics, and Social Science, as well as varying levels of teaching experience. 

\parabold{Interview Procedure}
Curators were interviewed via Zoom after completing at least one month of lesson plan curation, allowing them adequate time to engage in the curation process.
These interviews explored their curation processes, perceptions of lesson plan quality, cultural relevance, observed biases, and their understanding and views of AI. 

Teacher interviews were conducted two months after the Distribution Platform was deployed, allowing teachers time to interact with the tool for lesson plan creation. The interviews focused on teachers' experiences with the platform, usage of AI features by the power users, classroom integration, cultural appropriateness of LPs, and the implications on their broader work practices. 
A total of 17 teachers were interviewed in-person at the teachers' respective schools and five teachers were interviewed over Zoom.

Mentor interviews were conducted over Zoom two months after the deployment to gain insight into the teacher training process, teachers' usage of the tool and LPs, challenges encountered in classrooms, and the forms of support teachers required.

\parabold{Data Collection and Analysis}
The interviews were conducted in English or in Kannada with interpretation support, based on the participant preference. All interviews were recorded with participant consent, transcribed, and cleaned. Transcripts were translated where needed and de-identified for analysis. We used thematic analysis \cite{clarke2017thematic} to analyze the interview data and developed an initial codebook based on the research questions and interview protocols. Two researchers independently coded the transcripts and met twice weekly to add, review, and refine codes, resolve disagreements, and ensure coding consistency. The codebook was updated iteratively as new themes emerged from the data. The iterative discussions throughout data collection and coding led to consensus. Our final codebook (see Appendix \ref{codebook}) had 28 codes (e.g., catering to different student levels, reduction in workload, and lack of global context), clustered into five themes (e.g., quality of LPs, cultural relevance and bias, impact on teaching practices and AI knowledge and perception).

\subsection{Surveys}

We conducted structured pre- and post-intervention surveys to quantitatively evaluate the impact on teachers’ stress levels, access to instructional resources, time spent on work tasks, and trust in AI-based educational technologies. The same survey instrument was administered in both phases, with additional questions in the post-survey to learn about teachers' use of lesson plans from \toolname. The pre-survey was conducted prior to teacher training and tool deployment, while the post-survey was administered three months after the deployment of the tool.

\parabold{Teaching-Related Stress} Lesson planning is a major source of workload and stress for teachers \cite{mocklerOutsourcedCurriculumPlanning2024, riegCopingStressInvestigation2007}. To assess the tool’s impact on stress, we used the Teacher Stress Scale (TSS) developed by Chen et al. \cite{chenThrivingResilienceStress2022}, which includes two subscales: stress arising from teaching demands and stress related to limited school-based support. The former assesses pressures such as excessive workload, limited preparation time, and challenges in addressing diverse student needs; the latter captures perceptions of insufficient administrative and peer support. To align with other instruments, we adapted the original 5-point Likert scale to a 7-point format (1 = ``strongly disagree'' to 7 = ``strongly agree''). 
The subscales showed strong internal consistency in our data: teaching demands had $\alpha = .79$ (pre) and $.83$ (post), while school-based support had $\alpha = .70$ (pre) and $.83$ (post).


\parabold{Access to Teaching Resources} We developed a custom scale to assess whether the tool enhanced teachers’ access to contextualized instructional resources. Items used a 7-point Likert format to evaluate ease of access to interactive activities, culturally relevant LPs, digital content, and real-world curriculum-aligned examples. Pedagogical experts from the implementation partner reviewed items for content validity. The scale demonstrated strong internal consistency ($\alpha = .80$ pre, $.86$ post).

\parabold{Time Use} We measured weekly time spent on lesson planning and administrative tasks to evaluate workload shifts before and after tool adoption. In the post-survey, we added additional questions asking teachers to estimate the amount of time they spent preparing \toolname LPs to satisfy record-keeping requirements and for classroom delivery. The question design was grounded in the research team’s understanding of teacher workflows and refined with input from the implementation team.

\parabold{Trust in AI-Based Educational Technologies} We measured teachers’ trust in AI-based educational tools using an instrument adapted from Viberg et al. \cite{viberg2024trust}, based on Mayer et al.’s trust framework \cite{mayer1995integrative}. The scale covers three domains: perceived benefits (e.g., efficiency, support), perceived concerns (e.g., reliability, transparency), and willingness to adopt AI. 
The scale had excellent internal consistency in our sample ($\alpha = .89$ pre, $.91$ post).

\parabold{Survey Administration and Data Processing} The survey was administered via Qualtrics and shared with teachers through WhatsApp groups managed by mentors from the deployment team. Phone numbers served as identifiers to link pre- and post-intervention responses with lesson planning platform accounts. Participants were assured of confidentiality and informed that responses would not affect their professional standing. We received 601 complete responses for the pre-survey and 406 for the post-survey; analyses were conducted on the 406 matched responses. In terms of data cleaning, the missing values were imputed using \texttt{miceforest} \cite{MiceforestMultipleImputation, stekhoven2012missforest} and numerical outliers were winsorized \cite{SULLIVAN2021530} at the 95th percentile.  All quantitative analyses were performed using Python-based statistical libraries. Composite scores for each construct were calculated as the mean of corresponding items. Full survey instruments are included in Appendix \ref{survey}.
{\renewcommand{\arraystretch}{1.1}
\begin{table}[t]
  \caption{Demographic Profile of Teacher Participants}
  \label{tab:demographics}
  \begin{tabularx}{\textwidth}{@{}lX X@{}}
    \toprule
    \textbf{Item} & \textbf{Interview (n = 22)} & \textbf{Survey (n = 406)} \\
    \midrule
    Gender & Female: 68.1\%; Male: 31.9\% & Female: 56.4\%; Male: 43.6\% \\
    Platform Access Role & Standard: 54.5\%; Power: 45.5\% & Standard: 76.1\%; Power: 23.9\% \\
    Platform Usage Frequency & Less Active: 36.4\%; Active: 63.6\% & Less Active: 50.9\%; Active: 49.1\% \\
    Teaching Experience (years) & $\mu = 16.22$, $\sigma=5.83$, Range: 7 - 27 & $\mu = 13.98$, $\sigma=16$, Range: 1 - 32 \\
    Grades Taught & VI: 11; VII: 12; VIII: 14; IX: 9; X: 8 & VI: 234; VII: 241; VIII: 251; IX: 132; X: 143 \\
    Subjects Taught & Math: 10; Science: 11; English: 5; Social Sci.: 5 & Math: 170; Science: 140; English: 129; Social Sci.: 88 \\
    \bottomrule
  \end{tabularx}
\end{table}}
\subsection{Log Analysis}

\parabold{Curation Platform Logs}
Log data from the Curation Platform was used to evaluate the quality of AI-generated LPs and to examine how curators engaged with and refined the content. A total of 991 curated LPs (965 English, 26 Kannada) were analyzed. For each lesson plan, we had access to both the original AI-generated version and the corresponding edited version produced by the curators. We utilized Python to computationally identify and quantify edits at the content block level across the LPs. Quantitative analysis focused on the frequency and distribution of edits across content blocks to uncover curation patterns. Curator ratings for each content block were also analyzed descriptively to assess perceptions of content quality. In parallel, qualitative analysis was conducted using thematic analysis to categorize the types of edits made. We employed an inductive–deductive approach \cite{fuster2022inductive}, iteratively reviewing curator edits and comments to identify emergent themes and then grouping them into higher-order categories. The curators’ comments accompanying content block ratings were coded at the content block level, with codes capturing both evaluative judgments and rationales for revisions.


\parabold{Distribution Platform Logs}
Logs from the Distribution Platform were analyzed to understand how teachers interacted with curated LPs and AI-enabled components of the system. We examined a total of 5,544 LPs (4,064 English, 1,480 Kannada) created by the teachers using curated LPs. Although the curated Kannada-medium LPs were fewer than the English-medium ones (as seen in the previous section), they were more extensively utilized for generating LPs for classroom use, as all teachers taught in Kannada. Kannada-medium teachers also relied on English-medium lesson plans when a corresponding Kannada version was unavailable, as they were able to understand English. The edits by the teacher were computationally identified and analyzed using Python, following the same process as in the curation analysis. Quantitative analysis examined the number and frequency of edits per content block and per lesson plan. Thematic analysis was used to qualitatively examine the nature and instructional intent behind teacher modifications. Teacher ratings of LPs were also analyzed descriptively. Additionally, we analyzed teacher interactions with two chatbot features: \textit{Edu Chat}  and \textit{Lesson Chat}. Quantitative metrics included the number of users, interaction frequency, and message volume. Chat transcripts were thematically coded to capture recurrent patterns in teacher–AI interactions. We employed an inductive-deductive coding approach \cite{fuster2022inductive}: first, transcripts were read iteratively to identify recurring themes, and then these were organized into categories (e.g., assessment generation, multilingual support, lesson plan requests) as reflected in the codebooks (Appendix~\ref{chatlog-codebook}). Each transcript was segmented at the query level, and codes were assigned to individual queries rather than entire conversations.


\section{Findings}
Our findings revealed critical insights regarding how 
teachers generated and engaged with the AI-generated LPs. Specifically, we observed how 
teachers addressed gaps in AI-generated LPs, used the system to support both instructional and non-instructional tasks, assessed the relevance and usefulness of AI-generated learning content in classroom settings, and navigated the broader impact of the tool on teachers' work practices.  

\subsection{Collaborative Curation and Adaptation of AI-Generated Lesson Plans}
Curators and teachers adapted and augmented the LPs in distinct ways: curators prioritized content accuracy and general applicability across diverse teaching contexts, while teachers tailored the material to suit their specific classroom needs.

\subsubsection{Minor Edits and High Reliability in English Content.}

The English-medium LPs were reported by curators to be of high quality and suitable for classroom use. Out of 7,744 content blocks reviewed, 96.13\% were rated suitable for classroom use, 3.53\% required minor adjustments, and only 0.34\% were deemed unusable due to technical errors. These errors typically involved incomplete content generation or system-generated error messages and were addressed by regenerating the content. The quality of the LPs was closely tied to the quality of the Learning Objectives (LOs). Curators noted that the AI-generated LOs were often superficial and failed to capture the full breadth of the topic, requiring extensive edits. For example, in the \textit{Reflection of Light} lesson, a Science curator observed that the LOs were limited to basic definitions, omitting key concepts such as image formation and mirror types. Curators revised the LOs to align with Bloom’s Taxonomy \cite{forehand2010bloom}, adding key terms to enhance conceptual depth and ensure comprehensive curricular alignment (see Appendix Table~\ref{tab:curator_edit_comparison}). They found that strengthening the LOs often led to high-quality LPs that required minimal subsequent editing. Curator 5, who taught English and Social Science, and has curated 42 LPs, explained:

\begin{quote}
For identifying the learning outcomes, an expert teacher is definitely required. But my overall experience with the AI-generated content has been good. There were only a few minor changes that I had to make. Even a regular teacher while going through the content would be able to spot those two or three small things that need to be changed. So, while an expert teacher is very much needed for identifying learning outcomes, such expertise may not be as necessary for reviewing the LPs themselves.
\end{quote}


\parabold{Type of Edits}Overall, edits to the English-medium LPs were minimal, with only 314 out of 7,744 content blocks (4.1\%) requiring modifications. We categorized these edits into formatting edits (70.70\%) and content edits (29.30\%). Formatting edits involved minor adjustments such as line breaks, text styling (e.g., bold or italics), and punctuation corrections—none of which significantly altered the instructional content. Content edits addressed more substantive pedagogical issues, including clarifying explanations, ensuring curricular alignment, enhancing student engagement, and correcting factual inaccuracies. Assessment questions—particularly Multiple Choice Questions—were a common focus of content edits due to frequent inaccuracies in answer choices. Curators corrected these inaccuracies and annotated the correct answers clearly to aid teachers during classroom instruction. Curators exercised extra caution with assessment items due to their direct usage by students. Despite these focused corrections, the overall number of edits remained low.
\begin{figure}[t]
    \centering

    \begin{minipage}[t]{0.49\textwidth}
        \centering
        \includegraphics[width=\linewidth]{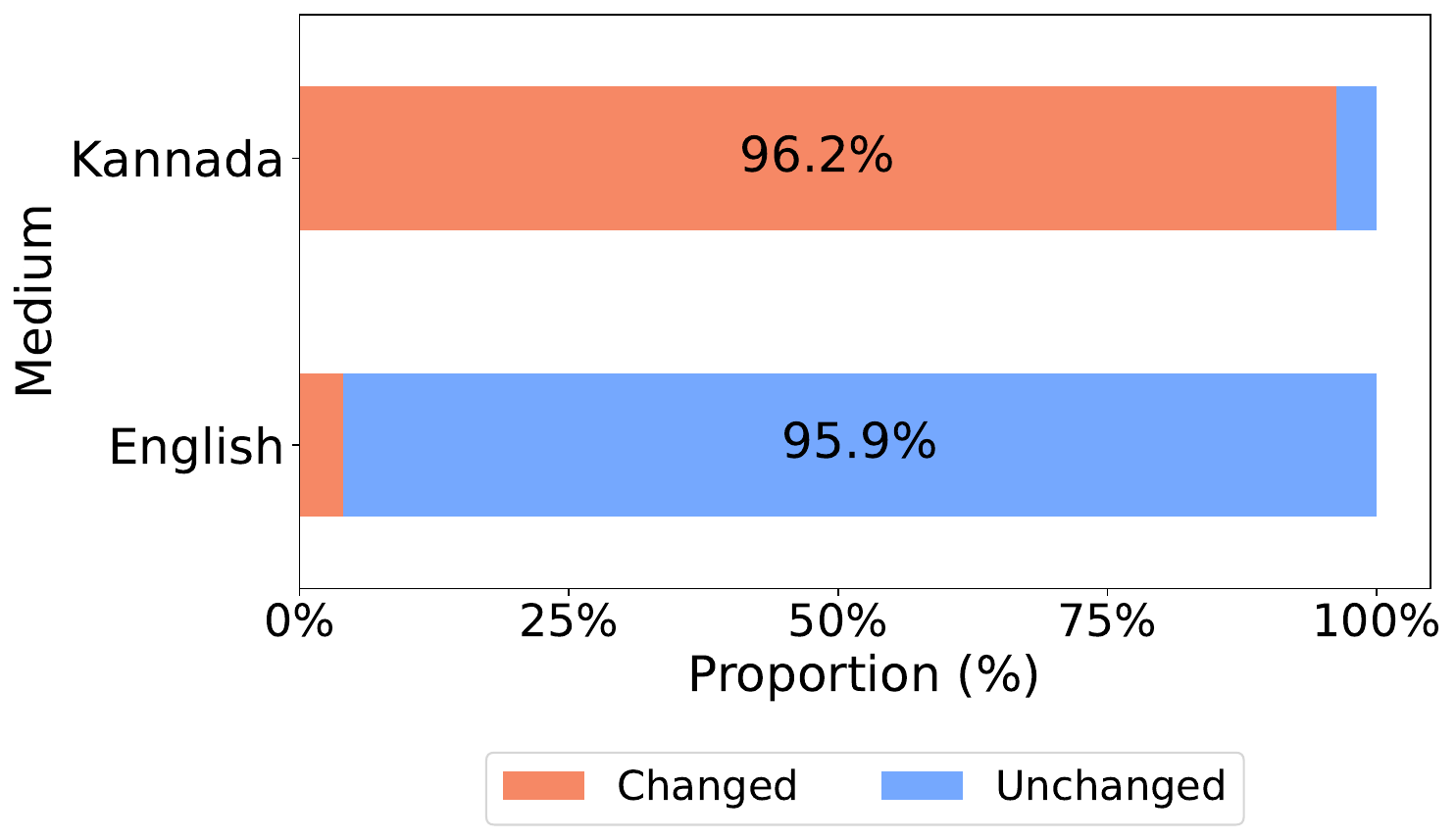}
        \caption{Curator Edits to Lesson Plans by Medium}
        \scriptsize{Proportion of content blocks that were changed versus unchanged during the curation of LPs, disaggregated by language medium.}
        \label{fig:CuratorChangesByMedium}
    \end{minipage}
    \hfill
    \begin{minipage}[t]{0.49\textwidth}
        \centering
        \includegraphics[width=\linewidth]{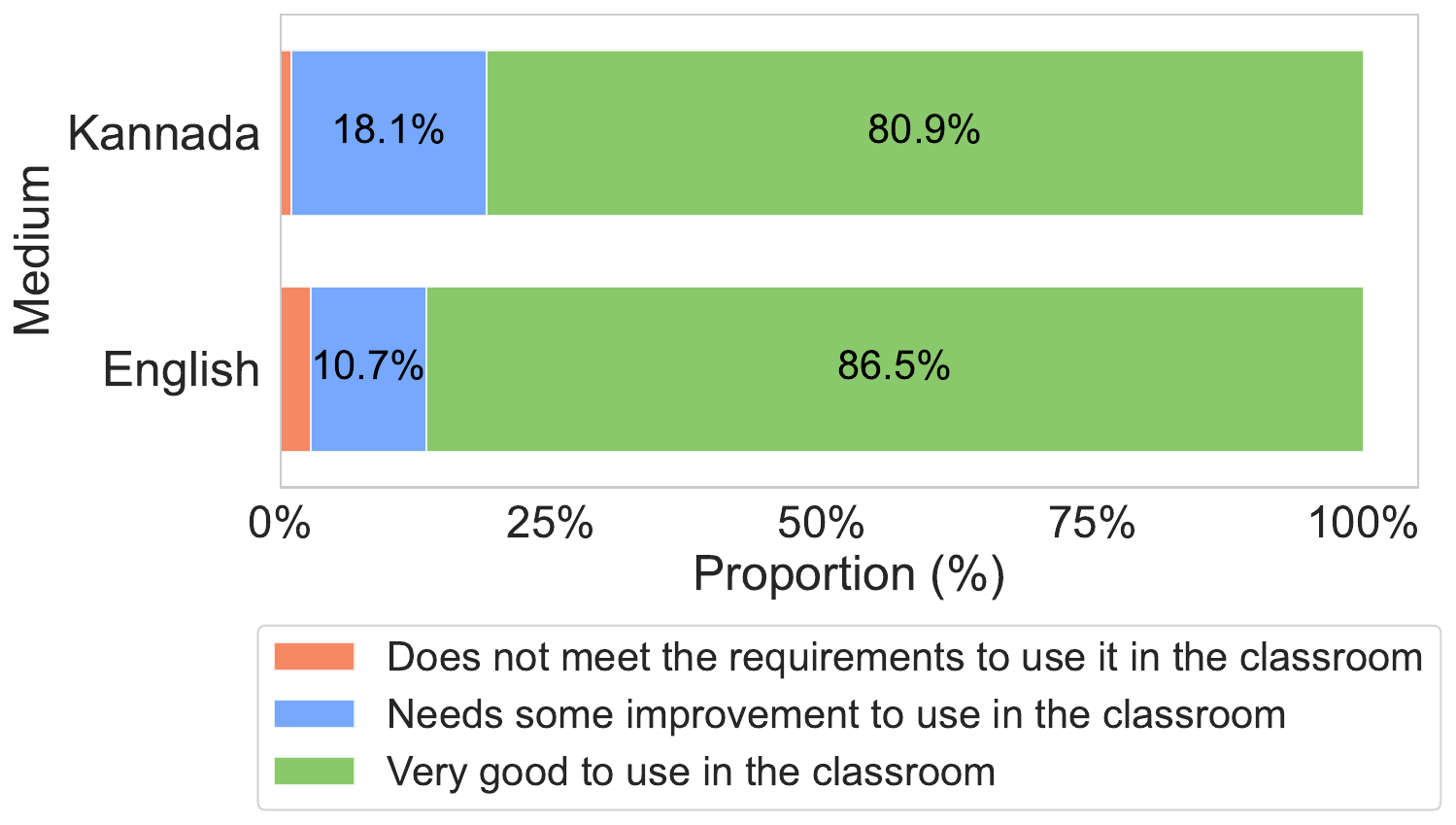}
        \caption{Teacher Rating for Lesson Plans by Medium}
        \scriptsize{Proportion of Kannada and English-medium lesson plans rated by teachers as “very good,” “needs improvement,” or “not usable” after generation for classroom usage.}
        \label{fig:TeacherFeedbackByMedium}
    \end{minipage}

\end{figure}

\parabold{Variation in Edits}There was notable variability in edit distribution by subject area and curator. Subject-specific edit rates ranged from a low of 0.82\% in Social Science to a high of 6.16\% in English, indicating moderate variability (variance = 6.19, coefficient of variation = 0.75).
English and Mathematics required more substantial edits compared to Social Science and Environmental Studies, where edits predominantly involved formatting. The Mathematics curators reported that most of their edits addressed errors in the assessment section, such as incorrect options or repeated answers in multiple-choice questions—a pattern that was also confirmed through content analysis of the edits. In English, by contrast, the edits were largely subjective, suggesting that the edit rate depended more on the curator’s discretion. Overall, variability among individual curators was greater than differences across subjects, with edit rates ranging from 0\% to 16.67\% ($\mu=4.87\%$, $\sigma=4.92$, variance = 24.22, CV = 1.01). A chi-square test confirmed significant differences among curators ($\chi^2$ = 272.39, df = 15, p < 0.001), though with a small effect size (Cramer’s V = 0.1768). Notably, the variance attributed to individual curators was nearly four times greater than subject-specific variance (variance ratio = 3.91), suggesting individual reviewing styles significantly shaped editing practices more than differences across subjects. We also analyzed the ratings which curators gave to the generated LPs, and the Shannon entropy scores for the ratings, 
ranging from 0 (highly uniform ratings) to 1.35 (diverse use of rating categories), reinforced our finding that curators exhibited varying assessment patterns. Our analysis of interview data also supported these quantitative and log analyses, 
revealing that some curators perceived themselves as more rigorous reviewers compared to others. Nevertheless, despite variability at the curator level, the overall extent of editing remained consistently low, underscoring the general reliability of AI-generated content.

\subsubsection{Linguistic Challenges in Kannada Content.}
While the edits in the English-medium content were minimal, curators made substantially more changes to the Kannada-medium LPs. Despite the pedagogical structure of the LPs being largely sound, curators identified widespread issues with the linguistic quality of the AI-translated Kannada content. Curator 14  from a Science background who reviewed Kannada LPs explained:

\begin{quote}
When the AI translated the English lesson plans to Kannada, it was done quite literally, but Kannada is quite different from English. In English, you might say, `Will you come?' but in Kannada, 
if you translate it literally, it doesn't make much sense.
\end{quote}

An analysis of 26 Kannada-medium LPs across three subjects revealed that 94.4\% of content blocks were edited. Only 2.5\% of the edited blocks involved formatting-only changes; nearly all edits were substantive and related to language. The extent of changes varied across subjects (Mathematics: 99.0\%, Science: 97.2\%, Social Science: 88.9\%), with the differences being statistically significant ($\chi^2$ = 10.25, $p$ = 0.0059). This suggests that while language quality issues were present across all subjects, certain domains—particularly Mathematics—required more extensive intervention. Subjects like Maths use specific academic terms---for example, \textit{guṇākāra} (multiplication) and \textit{samamiti} (symmetry)---in Kannada that are uncommon in everyday language, which likely led to more edits. In all subjects, the edits occurred at both the sentence and word levels.

\parabold{Nuanced Edits in Kannada}At the sentence level, curators observed that longer sentences often became grammatically incoherent. Literal translations from English frequently led to phrasing that was syntactically correct but semantically confusing or unnatural in Kannada. These sentence structures disrupted comprehension and required rewording to restore clarity. At the word level, the changes made were more nuanced and fell into several categories. First, curators replaced terms that were not typically used in Kannada-medium textbooks with standard equivalents. For example, \textit{svātantrya hōrāṭa} (freedom fight) was replaced with \textit{svātantrya saṅgrāma} (freedom struggle). Second, they corrected incorrect words that were slightly similar in meaning but inappropriate in context, such as \textit{guṇikānta} being replaced with \textit{guṇākāra} for multiplication, and \textit{samarūpate} (uniformity) revised to \textit{samamiti} (symmetry). Third, in cases where the meaning was accurate but not specific enough, curators substituted words that conveyed the intended meaning more precisely—for instance, \textit{kannaḍigaḷu} (glasses) was replaced with \textit{darpaṇagaḷu} (mirrors), and \textit{aṇṭugaḷu} (glues) with \textit{aṇṭisuva sāmagrigaḷu} (adhesive materials). Additional refinements were made to clarify vague or ambiguous terminology, such as replacing \textit{kālanukramaṇi} (chronology) with \textit{kālarekhē} (timeline), and \textit{vinimaya paddhati} (exchange system) with \textit{vastuvinimaya paddhati} (barter system).

Further edits involved stylistic changes to improve sentence flow, such as adjusting word order and tone to make explanations more fluent and accessible. In some cases, curators replaced code-mixed language constructions that combined English and Kannada with fully localized Kannada expressions. The translated content at times retained English terms unnecessarily or used them inappropriately within Kannada sentences, resulting in awkward phrasing and loss of clarity. For example, words like \textit{timeline}, \textit{mirror}, or \textit{glue} were used directly instead of their Kannada equivalents, and phrases such as \textit{oxygen cycle} or \textit{exchange system} appeared in hybrid forms like \textit{oxygen cakra} or \textit{vinimaya system}. Curators systematically replaced such instances with appropriate Kannada translations or transliterations and terminology to ensure alignment with textbook language conventions and improve accessibility for teachers. 

\subsubsection{Teachers' Flexible Use and Off-Platform Adaptation.}\label{subsub:off-platform-adaptation} Teachers regularly adapted the curated LPs during use, though they rarely recorded these edits on the platform itself. Of the 5,542 LPs that the teachers created from the curated base LPs, only 85 (1.53\%) were modified. A similar trend was seen with Kannada-medium plans, where just 19 out of 1,480 (1.28\%) showed edits. This low edit rate, however,  does not indicate passive acceptance or blind trust in the system’s outputs. Nor does it suggests that the teachers used the plans exactly as generated. 

Our interviews found a more nuanced picture of engagement. While newer teachers tended to follow written plans more closely, more experienced teachers often relied on mental planning—drawing from professional experience and a range of informal resources. Teachers viewed \toolname{}’s lesson plans as one of several resources they could access. Many adapted or extended the plans using materials from trainings, peer networks, cluster meetings, and WhatsApp groups. However, since these adaptations were largely done mentally, they were rarely recorded in the system—especially due to time constraints. Teacher 7, who's been teaching English for 18 years, shared:
\begin{quote}
The tool reduced our burden. Teachers don’t always have time to sit and write lesson plans because of other responsibilities. When we receive something already prepared, we can easily build on it by adding our own ideas. That’s a big help. ... In reality, we write and submit the same lesson plans each year just for records. Actually, we teach much more than what’s written down. A single lesson takes 10–15 class periods to teach, but we only submit a one-page plan. It’s not possible to capture everything we do in class in writing.
\end{quote}


In the next sections, we see in detail how teachers utilized their own resources and the AI-capabilities on the go to adapt \toolname's LPs as they used them to meet their instructional and non-instructional needs.

\subsection{How Teachers Used the AI System?}

Teachers used \toolname to support both instructional and non-instructional tasks. Based on usage patterns, we observed two groups: single-day users, who generated all LPs for the academic term (10 weeks) in one sitting, and multi-day users, who used the platform across multiple days. Interviews revealed that many teachers preferred to complete planning at once because they struggled to find time during regular school days. Single-day users made up 60.8\% (n = 622) of all users, generating an average of 2.62 LPs ($\mu$ = 2.62, $\sigma$ = 1.80; range = 1–13) in a single session. In contrast, multi-day users (39.2\%, n = 401) engaged with the tool across multiple days ($\mu$ = 3.3 days, $\sigma$ = 2.1; range = 2–14), producing more LPs overall ($\mu$ = 9.8, $\sigma$ = 12.4; range = 2–159).  Teachers used \toolname in three main ways: to complete record-keeping work efficiently, to support classroom teaching, and to create student assessments.

\subsubsection{Fulfilling Administrative Requirements Efficiently.}



All teachers were mandated by higher authorities to submit lesson plan documentation for record keeping—a responsibility they described as both significant and time-consuming. Prior to using \toolname, teachers typically spent 45–90 minutes preparing these plans. While peer networks and training materials offered LPs, these rarely followed the mandated 5E format, often forcing teachers to rewrite them. Limited digital access further compounded the burden: many lacked computers and digital proficiency, leading them to handwrite their plans, which increased their workload. \toolname eased this burden by generating submission-ready LPs in the required 5E format. Most teachers completed the task in under 15 minutes. Because the tool was accessible via smartphones—a device they were already comfortable using—it helped overcome both infrastructural and skill-related barriers. Log data corroborated these insights: 87.6\% of teachers accessed \toolname via smartphones, underscoring the importance of designing for mobile-first contexts.

While the record-keeping practices varied—some school authorities accepted printed copies, while others still required handwritten ones—teachers consistently described \toolname as time-saving. It allowed them to either print the plans directly or use them as a reference when writing by hand. Authorities who insisted on handwritten lesson plans had long justified the practice on the grounds that writing by hand ensured teachers actively engaged with the content, rather than merely reproducing plans circulated online or within teacher groups. This requirement predated the advent of AI and reflected broader concerns about teacher accountability. Despite these differing requirements, teachers found that the tool streamlined bureaucratic demands and freed time for actual classroom preparation. Teacher 3, who has been teaching Mathematics for 17 years, noted:

\begin{quote} It’s very nice. It saves a lot of time for teachers. Higher authorities keep asking for programs of work and lesson plans again and again. We have to keep writing them. Because of that, we don’t get time for actual class preparation. All the time goes into writing lesson plans. They always want to see the lesson plan. \end{quote}

Teachers also valued that the generated documents included their names, giving them a sense of ownership and agency. This feature helped them demonstrate to authorities that they were the creators of the submitted plans. 

\subsubsection{Supporting Effective Instructional Planning and Assessments.}
Teachers utilized the AI-generated LPs extensively to support their classroom teaching. Although teachers already had access to other instructional resources through training programs, WhatsApp groups, online forums, and previous teaching materials, they valued \toolname's LPs more because of the structured and detailed approach. The clearly articulated 5E framework significantly helped teachers organize instructional sequences effectively and provided a cohesive scaffold for lesson delivery.  Teacher 15, a Mathematics teacher with 20 years of experience outlined the benefits of using AI-generated LPs: 

\begin{quote}
It's staggered in such a way that it could be helpful for beginners and intermediate students. That is very helpful. The \toolname lesson plan gives us a detailed explanation. Teachers can read what to do in the classroom at each stage, in five stages. It's a very detailed lesson plan.
\end{quote}


Teachers appreciated that the LPs included a wide range of classroom activity ideas, clear explanations of complex concepts, and contextually relevant examples. For instance, an early-career Grade 7 Math teacher highlighted how the \textit{Geometry} lesson plan included relatable real-world examples, such as household objects and local buildings, making abstract concepts more accessible. Previously, teachers often relied on extensive external resources, including internet searches and multiple textbooks, to create engaging, activity-based lessons. \toolname significantly reduced this burden by providing comprehensive, ready-to-use content in one place. Teacher 18, a Science teacher with 10 years of teaching experience, explained:


\begin{quote}
We had to refer to the internet and books. For certain topics, we didn't get all the information because we had to refer to CBSE books or other sources. To make it more activity-based, even though it's in the textbook, for more ideas and concepts, we had to refer to other books. But it's easier now.
\end{quote}


\parabold{Chat assistance for customized resources}Beyond LPs, teachers actively used \toolname's chat features—Lesson Chat and Edu Chat—to quickly obtain supplemental teaching resources. Lesson Chat was used 111 times ($\mu = 2.52$ messages/chat, $\sigma = 2.43$), and Edu Chat was used 82 times ($\mu = 1.95$ messages/chat, $\sigma = 1.64$). Although Lesson Chat was intended for lesson-specific questions and Edu Chat for general educational queries, teachers often did not differentiate their use distinctly. Queries primarily focused on obtaining assessment questions for both formative and summative \cite{black1993formative} purposes, such as \textit{unit test papers}, \textit{one-mark questions for science chapters}, and \textit{objective-type questions on probability}. Additionally, teachers requested engaging formative assessment activities, including \textit{puzzles or riddles for algebraic expressions}, \textit{questions for agree and disagree activities for integers}, and \textit{objective questions for slap the board activities}. One teacher noted the efficiency gained by generating questions for the \textit{Slap the Board} activity, which typically required over an hour but was completed in less than five minutes using \toolname. Teachers also sought practical classroom resources like \textit{phonics activities for AEIOU sounds}, \textit{paper folding activities}, and contextual information such as \textit{examples of crops in Karnataka}, \textit{uses of forests in Kannada}, and \textit{details about the Taj Mahal in Kannada}.

Another highly valued feature was Question Gen, introduced during the final month of the intervention following teacher requests. Teachers quickly adopted this feature, generating 211 customized question papers across beginner, intermediate, and advanced levels. Teachers particularly appreciated \toolname's original and concept-teaching questions, which were otherwise challenging and time-consuming to create. These assessments encouraged students to think critically rather than rely solely on memorization. Survey findings further confirmed that assessment resources were among the most frequently utilized resources of \toolname.

\subsection{How Effective Were AI-Generated Lesson Plans in the Classroom?}

Teachers expressed high satisfaction with \toolname's LPs: roughly 85\% of plans were marked \textit{Very Good}, 13\% as \textit{Needs Some Improvement}, and only 2\% as \textit{Does Not Meet Requirements}. However, as shown in Figure \ref{fig:TeacherFeedbackByMedium}, we observed notable differences across language mediums. English-medium plans received more \textit{Very Good} ratings (87\%) and fewer \textit{Needs Improvement} ratings (11\%) compared to Kannada-medium plans (81\% and 18\%, respectively). These differences were statistically significant ($\chi^2$ = 66.79, p < 0.001, Cramer’s V = 0.110). Still, both mediums had very low rejection rates---roughly 3\% for English and 1\% for Kannada---indicating a high baseline quality. While these quantitative results point to broad satisfaction, interviews provided further insight into the usability, contextual alignment, and pedagogical value of the AI-generated content.

\subsubsection{Flexible Content with Pedagogical Breadth, but Repetitive Patterns.}
Even though not all content in the LPs was directly applicable to every classroom, teachers appreciated the breadth and flexibility the plans offered. The inclusion of a wide range of examples, explanations, and activities allowed them to select those that best matched their teaching styles, student needs, and classroom contexts. Teachers valued having multiple options, which enabled them to tailor their instruction to different learning levels. 
This flexibility reduced the need to consult external resources and helped teachers deliver lessons more efficiently. Teacher 17, an English teacher with 21 years of teaching experience, described:

\begin{quote} There are different activities for different types of learners. Some people may be good in speaking in the classroom, some people may be good when they're in a group. So there are activities for different kinds of learners is what I felt. \end{quote}


 While AI-generated LPs featured diverse topics and pedagogical approaches, activity types—such as treasure hunts and matching exercises—frequently repeated across LPs. This repetition primarily stemmed from deliberate system design choices aimed at practicality in low-resource classrooms, where digital infrastructure and teaching materials are often limited or unavailable. Consequently, \toolname's LPs were purposefully created to be usable and engaging without additional infrastructural or material demands (covered in more detail in \S\ref{activity-section}). Although some curators perceived the repetition of activities as a limitation, teachers generally did not view it negatively; instead, they felt comfortable modifying activities and adding their own twist to better engage their students. Additionally, while the current system permitted teachers to edit only their individual copies of LPs, many expressed interest in sharing their classroom-tested activities to a collective resource pool, enriching available content and supporting the broader teaching community.

\subsubsection{Lacking Visual Support.}

Teachers frequently identified the absence of visual aids—particularly images and videos—as a significant limitation. Although this was partly due to the intentional design choice to keep LPs independent of digital resources, the issue extended further: even the lesson plan documents themselves lacked visuals, largely because \toolname's underlying LLM system is text-focused. Teachers described the LPs as excessively text-heavy, offering no visual support for both their own comprehension and student learning. This gap was especially problematic in subjects like Science and Mathematics, where visuals are crucial for effectively conveying concepts. For example, one Math teacher found a lesson plan on \textit{parallelograms} challenging due to the absence of illustrative images, while a Science teacher noted similar difficulties with process-oriented topics like \textit{photosynthesis} and the \textit{life cycle of a cell}. Teachers noted that even before being introduced to \toolname, they regularly used their personal smartphones---due to the lack of digital infrastructure such as projectors---to show concept-related images to students to aid comprehension. While \toolname reduced preparation time overall, the complete absence of visuals in the LPs still required them to spend additional time searching online for suitable resources.

This limitation extended to the accompanying PowerPoint presentations, which similarly lacked engaging visual content. Although some teachers attempted to supplement these materials with online videos, they often found freely available resources unsuitable—either due to advertisements, overly complex language, or misalignment with curricular objectives. Teachers emphasized the critical need for concise, well-designed videos tailored to specific curricular topics. Additionally, they noted that high-quality visual resources, often created by teachers themselves, were already available through government channels but not integrated into the platform. Teachers suggested embedding such vetted resources directly into the system and expressed interest in contributing their own visual materials, which would significantly streamline preparation and benefit peers facing similar classroom challenges. 

\subsubsection{Strong Regional Relevance, Limited Relevance to Students' Competence.}
Teachers appreciated that the LPs included content at varying levels of difficulty (beginner, intermediate, \& advanced) which helped address the diverse learning needs within their classrooms. However, they also pointed out that these levels did not always align with their students’ actual competencies. In several instances, materials labeled as \textit{beginner} still assumed foundational skills that many students had not yet developed. For example, a Grade 6 math teacher noted that the beginner-level materials required numeracy skills that their students were still acquiring. This mismatch reflects a broader limitation of AI-generated content: the system lacks access to real-time, classroom-specific learning data and therefore cannot fully adapt to the range of abilities present in different student groups. Teachers emphasized that even experienced educators often struggle to assess student levels without regular classroom interaction, underscoring the difficulty of pre-designing instructionally appropriate materials. According to Teacher 10,  a Mathematics teacher, lesson planning is deeply situated, relational work that resists automation:

\begin{quote} I would trust AI for geographical relevance, but not to deliver effective lessons in my classroom because it doesn’t know the background of my students. \end{quote}

Teachers valued that the LPs incorporated culturally and geographically relevant examples drawn from their own state. These references sparked student curiosity, made abstract ideas more relatable, and increased engagement. A science teacher shared how a science lesson explained the process of rusting through the example of coastal Karnataka, where salty air accelerates corrosion—making the concept both concrete and locally meaningful. Similarly, a Social Studies teacher appreciated how historical examples were connected to familiar places in Karnataka, helping students connect the material to their lived experiences. Teacher 11, a Social Studies teacher, highlighted how difficult it is to design LPs that reflect regional diversity and found the tool helpful in doing some of that work:

\begin{quote} It is difficult when you're making a lesson plan to think about all the things happening in Karnataka - which region has what... So in that way, this was very helpful that the lesson plans have already given that information for different parts of Karnataka. \end{quote}

\parabold{Difficulties with local contextualization}At the same time, teachers expressed mixed views on the depth of contextualization. While state-level references were seen as valuable, some noted that examples from unfamiliar regions of Karnataka were not always meaningful to their students. Others felt the content lacked more immediate, hyper-local references tied to students’ neighborhoods or communities. This limitation in achieving hyper-local contextualization stemmed from the design of the RAG system, which generated content intended to cover the entire state at scale. Although teachers had both UI- and AI-based affordances to update content for their local context, they more often made mental adaptations off-platform to suit their students’ needs, as discussed in \S\ref{subsub:off-platform-adaptation}.  A few teachers also suggested that the focus on state-level content came at the expense of national or global perspectives. Still, these concerns were seen relatively as minor. Most teachers felt confident in filling such contextual gaps themselves and emphasized that even partial regional grounding was a helpful starting point. These findings underscore that contextualization is not a one-time design task but an ongoing, situated process where teachers engage with their own expertise and AI outputs.

\subsection{Broader Impact on Teachers’ Work Practices}
Our analysis found that \toolname significantly shaped teachers’ work practices on multiple dimensions.

\subsubsection{Less Time and Stress in Lesson Planning.}
Teachers reported that \toolname significantly improved their productivity by reducing the effort required for lesson plan documentation, streamlining planning tasks, and offering greater flexibility in when and how they worked. Time previously spent handwriting plans in the required format was largely recovered, as AI-generated plans could be used with minimal edits to satisfy administrative requirements.
The tool also gave teachers quick access to editable, high-quality instructional content—including explanations, classroom activities, and assessment questions—replacing the often time-intensive process of searching for or creating resources from scratch.

\parabold{Increased convenience}The smartphone-based design of the tool introduced a level of flexibility that was previously unavailable. While teachers had earlier prepared LPs at home after school hours—often at the cost of personal or family time—they could now plan during commutes or in other spare moments throughout the work day. This shift allowed them to better manage their workload without encroaching on personal time. The convenience and support offered by the tool stood in stark contrast to previous educational interventions, which were often seen as adding to teachers’ burden. Teacher 8, a Science teacher, with 20 years of teaching experience explained:

\begin{quote} When I was called for the \toolname training, I was like, oh no, one more headache. But really, I came back with a very nice experience. I felt like, this is a good one. Usually, whenever someone comes with something new [intervention], it just adds to our workload, but this is a burden-releasing concept. We already have so much on our plate, so anything extra just feels like more pressure. Especially in big schools, managing students itself is tough, and then training means more work. But this time, when I walked out, I genuinely felt that this one’s actually good. \end{quote}

Our survey results supported these qualitative findings. On average, teachers saved 2.02 hours per week on lesson planning ($\sigma = 4.1$), with more active users saving 2.64 hours per week ($\sigma = 4.0$) and less active users saving 1.50 hours per week ($\sigma = 4.0$). These reductions showed a small to medium effect overall ($d = -0.494$, $r = 0.336$), with a medium effect for active users ($d = -0.658$, $r = 0.415$) and a small effect for less active users ($d = -0.371$, $r = 0.262$). 

\begin{figure}[t]
    \centering
    \includegraphics[width=\linewidth]{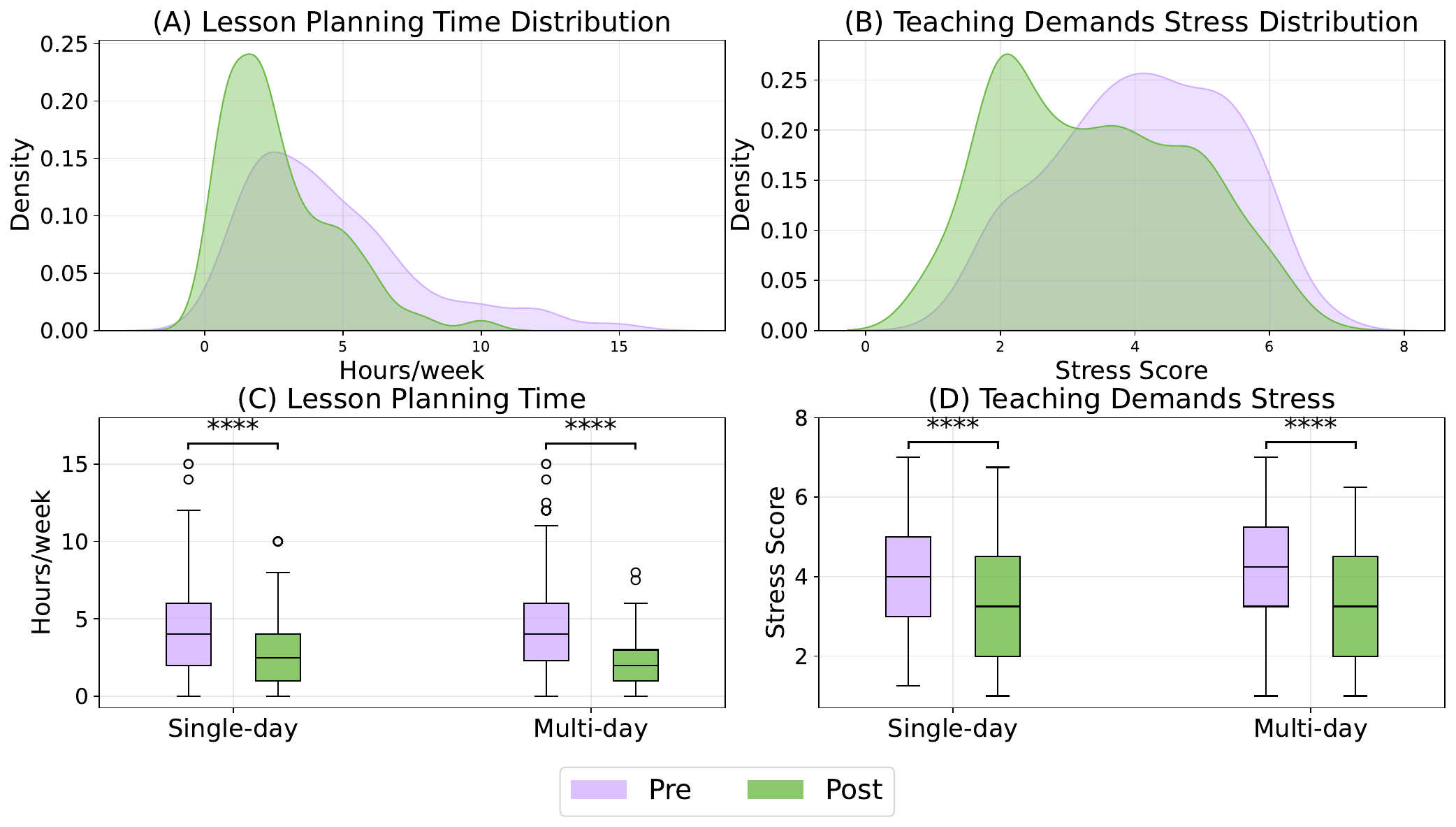}
    \caption{Lesson Planning Time and Teaching Demand Stress}
    \scriptsize{ Density plots (A, B) and box plots (C, D) for lesson planning time and teaching demands-related stress, comparing pre- and post-intervention values across Single-day and Multi-day user groups. Statistical significance levels are reported as follows: $p < 0.000005$ (****), $p < 0.00005$ (***), $p < 0.0005$ (**), and $p < 0.005$ (*). Shaded areas in density plots represent kernel density estimates for pre and post values.}  
    \label{fig:lp_time_stress_box_density}
\end{figure}

\parabold{Reduced stress}Teachers also reported a measurable reduction in teaching-related stress. On a 7-point Likert scale, the average \textit{Teaching demands related stress} score declined by 0.72 points ($\sigma= 1.66$; d = -0.436; r = 0.289), indicating a small effect size. For example, teachers who previously responded \textit{Somewhat Agree} to statements like \textit{"I felt stressed for not having enough time to complete my teaching work (e.g., preparing, teaching the curricular content)"} shifted closer to a \textit{Neutral} response by the end of the intervention. As shown in the Figure ~\ref{fig:lp_time_stress_box_density}, the post-intervention stress distribution not only shifted leftward but also became more concentrated in the lower stress range. Median scores decreased from 4.25 to 3.25. The effect was more pronounced among multi-day users, who showed an average decline of 0.85 points (d = -0.503), compared to a 0.60-point drop among single-day users (d = -0.371). These findings suggest that even modest reductions in stress were consistently experienced across user groups, with visible shifts away from moderate-to-high stress responses.


Together, these results suggest that the tool helped reduce both logistical workload and the cognitive stress of lesson preparation. However, despite the tool’s usefulness, both teachers and mentors emphasized that its effectiveness in improving student learning ultimately depends on how well teachers are able to integrate and use these LPs in practice.  

\subsubsection{Shift Toward Activity-Based Pedagogy.}\label{activity-section}

The adoption of \toolname signaled a shift toward activity-based teaching. Teachers had been using the 5E lesson plan format even before the introduction of \toolname, and they already recognized the value of classroom activities for fostering student engagement and deeper learning. However, they often struggled to design and implement such activities consistently due to limited time and a lack of readily available resources. \toolname partially addressed this gap by providing easily accessible, contextually relevant activities tailored to each topic. These activities—such as \textit{visiting a gram panchayat office} or \textit{collecting leaves with different venations}—were grounded in the local environment and did not rely on digital infrastructure. Unlike device-dependent solutions such as Kahoot \cite{KahootLearningGames} or Khanmigo \cite{KhanmigoFreeAIpowered}, the tool offered no-tech, practical ideas using commonly available materials, making them particularly well-suited for low-resource classrooms. One English teacher noted that the \textit{Clap and Tap} activity—where students clap upon hearing consonant sounds and tap their heads for vowel sounds—was both engaging and required no materials. Similarly, physical activities like \textit{treasure hunts} and \textit{relay races}—where students search for concept-related clues around the school while competing with peers—supported embodied learning by combining cognitive and physical stimulation. Teachers emphasized that the availability of such ready-to-use activities reduced their lesson planning burden and motivated them to experiment with new instructional approaches. Teacher 15, a Social Science teacher, explained:

\begin{quote} There is a change in the teaching style mainly because earlier teaching was just one way where we would just talk to the students. In my subject [Social Science], I was only verbally informing the students about each chapter, it could get boring for them. But now, for every specific topic, there are a lot of activities mentioned. I feel there's more involvement from the students through this activity-based teaching now. \end{quote}

The survey data corroborated these observations. In response to the question, \textit{“Overall, how many of the following resources from \toolname did you implement in your classroom teaching?”}, teachers reported frequent use of 5E instructional steps ($\mu = 6.03$, $\sigma = 6.87$) and classroom activities ($\mu = 5.99$, $\sigma = 6.06$) most frequently after assessment questions. Concept explanations ($\mu = 5.59$, $\sigma = 6.75$) and real-world scenario examples ($\mu = 4.60$, $\sigma = 5.31$) were also widely used, though to a slightly lesser extent (See Appendix Figure~\ref{fig:resource-usage}). While assessment resources met immediate institutional demand of conducting frequent student assessments, the high usage of activity-based and explanatory content reflected teachers’ intent to promote deeper engagement and participatory learning. Mentors also observed that the use of such activities was becoming more common in classrooms following the adoption of the tool.

\parabold{Contingent on systemic factors}While \toolname significantly reduced the time required for lesson planning, teachers emphasized that it had no impact on administrative tasks unrelated to teaching. Routine responsibilities like data entry and reporting continued to consume a large portion of their time. Even with easier access to high-quality, ready-to-use materials, many struggled to consistently implement activity-based teaching because administrative work often spilled into classroom hours—cutting into instructional time. One teacher pointed to the introduction of \textit{APAAR}, a government initiative requiring detailed student ID data entry, as a recent example of added workload. In the absence of adequate clerical support, such tasks routinely fall to teachers. While the potential of \toolname to support more interactive and engaging pedagogy was widely recognized, teachers emphasized that its long-term success depends on systemic support—particularly reducing non-instructional burdens and improving staffing. Teacher 14, a Science and Mathematics teacher, succinctly put it:

\begin{quote} There is no challenge for teachers in implementing these LPs. Teacher shortage is the main problem. Now we have only three teachers. One of our H.M. [Headmaster] went for training. Only three teachers, we handle seven classes. That is a government problem. \end{quote}




\subsubsection{Growing Awareness and Practical Engagement with AI.}

The intervention fostered greater awareness among teachers about the use of AI tools to support their teaching practices. While both pre- and post-intervention surveys reflected consistently high levels of trust in AI (Pre: $\mu=5.41$, $\sigma=0.80$; Post: $\mu=5.45$, $\sigma=0.85$), with no statistically significant change observed (Wilcoxon signed-rank test: $W=30565.50$, $p=0.3298$), this continuity in trust coincided with a marked increase in actual engagement with AI tools. Before the intervention, over 43\% of teachers reported to have never used ChatGPT, and 41.87\% had never used Meta AI. These figures dropped sharply after the intervention—to 11.33\% for ChatGPT and 10.34\% for Meta AI. Among the 119 teachers who had never used either tool before the intervention, 107 (89.92\%) reported using at least one of them afterward. Mentor 3, who supported 250 teachers across 187 schools in implementing \toolname, shared: 
\begin{quote} When we called them [teachers] for training, out of 20 people, only two knew about AI. Now, it's better, and they know more about AI. After using this portal, they have a much better understanding of AI. \end{quote}

Teachers who actively used \toolname also expressed a preference for its integrated chatbot features over standalone tools. Unlike generic AI platforms, \toolname's chat interface was directly linked to the lesson content, enabling teachers to ask questions without needing to re-explain the topic. Teacher 8, who teaches science, pointed out: 
\begin{quote} I ask questions to Lesson Chat. I can ask the same to MetaAI, but then I have to explain the lesson to MetaAI. Here I can ask directly. \end{quote}


\parabold{Shift in perception of AI}The intervention also contributed to shifting teachers’ perceptions of AI, particularly around its ability to handle context-sensitive and local-language inputs. A senior teacher with 25 years of experience admitted that they initially associated AI with \textit{"fake videos and images,"} but after using \toolname, came to see it as a practical and accessible tool for quickly retrieving relevant information. There was initial skepticism about whether AI could generate content that resonated with students, especially in regional languages like Kannada. One teacher shared that they had assumed such technologies would not function well in Kannada but were pleasantly surprised by \toolname's ability to interpret and respond appropriately to local-language inputs. These experiences suggest that while baseline trust in AI was relatively high, hands-on interaction with \toolname helped demystify its capabilities and enhance its perceived usefulness in everyday teaching. At the same time, teachers expressed concerns about overreliance. One experienced teacher joked that she worried she might \textit{“stop thinking about the topics”} because the AI provided information so quickly and easily. Others emphasized the importance of learning to use AI tools thoughtfully—not only to stay current alongside their students but also to exercise professional judgment to ensure these technologies are used in pedagogically beneficial ways.

\section{Discussion}
Our finding show that teachers engaged with \toolname not as passive users of AI, but as active collaborators who evaluated, adapted, and repurposed materials within the constraints of their classrooms and institutional demands. 
We build on prior work in teacher–AI collaboration, communities of practice, and the design of LLM systems for low-resource languages to examine how such tools are taken up in real-world settings---where language, infrastructure, and systemic pressures shape both the promise and limits of Human–AI cooperation.

\subsection{Considerations for Effective Teacher-AI Collaboration}
Teacher-AI Collaboration (TAC) is an emerging area of research that explores how teachers engage with AI tools in practice. Researchers have observed that this collaboration often develops in a sequence over time: starting with teachers as passive recipients of AI-generated content, then moving to more active use of AI tools, and eventually evolving into a partnership where teachers and AI work together as constructive partners \cite{kim2024}. In our intervention, we found that most teachers began as passive recipients, with some gradually becoming more active users. However, being a passive recipient did not imply uncritical acceptance of AI outputs. Teachers exercised professional judgment, selectively using content they found useful and adapting it with their own expertise. They drew on their own resources and expertise to modify or supplement AI outputs as needed. This aligns with prior research showing that teachers, as the primary agents in instructional design, draw on various forms of professional capital—including human, social, and decisional capital \cite{Liu_Li_2022}—to ensure that content aligns with their students’ needs and classroom contexts \cite{Cukurova_Kent_etal._2019}. We also found that the curriculum-integrated chatbot interface enabled a more reciprocal interaction between teachers and the AI, contrasting with the one-directional experience of using general-purpose tools like ChatGPT or Meta AI. In line with prior work that emphasizes the importance of such interfaces in facilitating meaningful teacher-AI interaction \cite{Leeuwen_Campen_etal._2021},  our findings suggest that this curriculum alignment was a critical factor in teachers’ preference for \toolname's integrated chatbot over standalone alternatives.

Our study found that effective AI integration was often impeded by challenges related to infrastructure, workload, content compatibility, and linguistic or cultural mismatches. Filiz et al. \cite{Filiz_Kaya_etal._2025} similarly identify key pedagogical enablers and barriers—such as content variety, interactive materials, and ease of content creation—which further support our observations. While prior research emphasizes the importance of Teacher AI Competency or AI Literacy—including not just technical proficiency but also ethical awareness and critical digital skills—as central to meaningful AI use in education \cite{Iryna_2025, Long_Magerko_2020, Ng_Leung_etal._2021}, our study did not focus explicitly on AI literacy. Future work could explore how varying levels of AI literacy shape tool integration, particularly in Global South contexts.

Further, while pedagogical and infrastructural factors shape the integration of AI tools, they do not fully explain what motivates teachers to adopt such tools in the first place. A central driver of \toolname's adoption was its ability to reduce the time and effort required for lesson plan record-keeping—a task that often imposes a significant bureaucratic burden on teachers we worked with. According to the Extended Unified Theory of Acceptance and Use of Technology \cite{venkateshConsumerAcceptanceUse2012}, \textit{hedonic motivation}—the pleasure derived from using a technology—can significantly influence behavioral intention. Although teachers did not describe lesson plan generation as enjoyable in itself, they appreciated the relief from the cognitive and mechanical labor associated with preparing plans manually. Given that workload from administrative record-keeping is a known stressor for teachers \cite{Davidson_2009,Hundani_Toquero_2021}, this indirect form of pleasure contributed meaningfully to technology adoption.

These findings highlight that the design of teacher-facing AI tools must go beyond instructional considerations and address the broader scope of teachers' responsibilities. In our context, tensions between teachers and bureaucratic mandates surfaced as a key site of friction. Aligning the expectations of multiple stakeholders—teachers, designers, and policymakers—may therefore be critical for fostering sustainable and effective Teacher-AI collaboration.

\subsection{Leveraging the Communities of Practice}

The concept of Communities of Practice (CoPs), introduced by Lave and Wenger \cite{lave1991situating}, offers a valuable lens for understanding how teaching is socially constructed through shared activity and meaning-making. CoPs emerge through \textit{mutual engagement}, \textit{a joint enterprise}, and \textit{a shared repertoire} of practices and resources developed through sustained participation \cite{wenger1999communities}. In education, CoPs have been shown to support pedagogical innovation, deepen teacher expertise, and foster sustained teacher learning and growth \cite{supovitz2002developing, paltiwaleUseCommunityPractice2020}. 

In India, this framework is particularly relevant given the scale and structure of the teaching profession. With nearly 10 million teachers across 1.47 million schools, the system is supported by over 600 District Institutes of Education and Training (DIETs), more than 6,000 Block Resource Centres (BRCs), and nearly 70,000 Cluster Resource Centres (CRCs) \cite{kaur_resource_centers}. These institutions form a hierarchical but collaborative support structure: DIETs coordinate district-level training, BRCs manage block-level programs, and CRCs facilitate localized mentoring and peer learning through regular cluster meetings \cite{wolfenden2015tess}. However, despite this extensive infrastructure, professional development often remains uneven in quality and disconnected from teachers’ everyday realities \cite{dyer2004knowledge}.

Our findings point to teachers' desires to embed AI-based tools like \toolname within these existing professional networks. Teachers were not only interested in using AI-generated LPs but also in sharing them with peers—framing lesson planning as a collective, not solitary, practice. \toolname could serve as a shared digital scaffold within this ecosystem—not just for teachers but also for specialists affiliated with DIETs, BRCs, and CRCs. Involving these stakeholders in curating, refining, and adapting content would foster collective ownership and transform the platform into a living repository of localized pedagogical knowledge. Personnel at resource centers, drawing on their contextual and linguistic expertise, could help ensure that AI-generated content reflects regional languages, cultural nuances, and student needs.

Embedding \toolname in this way is not only important for improving instruction but also for supporting teachers’ professional well-being. Prior research has shown that technology can reconfigure teachers’ work in ways that generate friction across institutional expectations \cite{varanasiHowTeachersIndia2019, chandranTeacherAccountabilityEducation2022}. In our study, teachers emphasized that for the tool to be sustainable, its outputs must be recognized by government officers for official record-keeping. Aligning AI-generated materials with the expectations of multiple actors—teachers, trainers, and administrators alike—is essential to securing long-term buy-in and reducing uncertainty.

This approach is consistent with prior research on effective teacher development. Studies from other educational systems show that when technology is embedded within strong professional communities, it is more likely to support sustained pedagogical change. For example, the Teaching and Learning International Survey (TALIS) 2013 found that teachers engaged in collaborative professional learning reported greater use of innovative pedagogies, such as small-group instruction, alongside higher job satisfaction and self-efficacy \cite{jerrim2019teaching}. In countries with high educational performance such as Finland, teacher collaboration is deeply embedded within professional culture and is closely linked to strong learning outcomes \cite{vangriekenTeacherCollaborationSystematic2015}. More broadly, research shows that participation in professional learning communities leads to improved student achievement, enhanced instructional practices, greater teacher self-efficacy, and higher job satisfaction \cite{christensen2025professional}. 
While lesson plans in \toolname were co-created through collaboration between teachers and AI, the tool fell short in enabling collaboration among teachers and other stakeholders within their Communities of Practice. Embedding AI tools like \toolname within these collaborative structures—rather than treating them as standalone solutions—can help nurture and amplify the existing strengths of professional communities, making educational innovations more contextually relevant, adaptable, and sustainable.

\subsection{Addressing the Shortcomings of LLMs for \lq{}Low-resource\rq{} Languages}

Prior work has documented a tendency among users to accept LLM-generated content without critical evaluation \cite{ramjeeASHABotLLMPoweredChatbot2025, mackoSpeculationMeasuringGrowing2025}, emphasizing the need for careful human review rather than superficial validation. \toolname adopted an human-in-the-loop (HIL) approach wherein curators reviewed and corrected the AI-generated LPs before they were made available to teachers. This strategy allowed teachers to engage with vetted content without needing to spend additional time identifying and rectifying errors themselves. This approach aligns with the broader paradigm of HIL systems, where data-driven AI outputs are enhanced by human contextual understanding \cite{natarajanHumanintheloopAIintheloopAutomate2025, wuSurveyHumanintheloopMachine2022}. Our study contributes to this line of work by shedding light on the dynamics of human contributions in educational HIL systems. In \toolname, each lesson plan was reviewed by a single curator. While this ensured a baseline level of quality, our findings revealed that the extent and nature of edits varied significantly across curators, echoing prior research showing consistent variance in human performance even when individuals engage in identical tasks \cite{jrMethodsIncludingHuman2022, viswesvaranPerspectivesModelsJob2000}.


The HIL approach proved especially valuable for Kannada-medium LPs, where curators had to modify over 95\% of the AI-generated content blocks.  
In our project, the learning content was first developed in English and then translated into Kannada. While this practical choice addressed immediate needs, it also risked reinforcing epistemic biases, as translation alone does not capture the distinctive knowledge systems embodied in each language \cite{camara-leretLanguageExtinctionTriggers2021, harrison2007languages, ExtinctionsLanguageDeathIntangibleCulturalHeritageandEarly21stCenturyRenewalEfforts}. Scholars have emphasized that each language functions as a vessel of unique worldviews and epistemologies, and the loss or underrepresentation of a language risks erasing valuable ways of knowing. Recent research has also highlighted that unified multilingual language models often fail to capture the diverse worldviews embedded in different linguistic traditions, tending instead to reproduce knowledge structures aligned with dominant languages such as English \cite{helmDiversityLanguageTechnology2024}. Rather than assuming that a single multilingual model can equitably represent all languages, there is a growing call to rethink model architectures, evaluation frameworks, and training objectives to better support linguistic and cultural plurality through the development of multiple LLMs in different languages \cite{clausNowYouAre2024}. Furthermore, the inability of current LLMs to produce high-quality, technically deep content in Kannada reflects model limitations, not an inherent incapacity of the language itself. Knowledge artifacts with technical depth can certainly be created natively in Kannada by proficient individuals. This requires systems that empower local knowledge holders to digitally share their expertise. In educational contexts, where the fundamental goal is to build collective knowledge and nurture future scholar communities, creating avenues for native knowledge production is especially crucial. Such an approach not only addresses immediate educational needs but could also serve as a foundation for developing more robust, truly representative language models. Rather than relying solely on translations from English, the creation and aggregation of native content could support the growth of LLMs that authentically embody the epistemologies and intellectual traditions of underrepresented languages. 

To further address the shortcomings of LLM-generated content for so called \lq{}low-resource\rq{} languages, we argue that crowd-sourcing mechanisms can play a vital role in two complementary ways. First, crowd-sourcing can support mistake correction: when a user corrects an error, that correction can propagate to benefit all future users. Second, crowd-sourcing can enable the contribution of new ideas, where users share original knowledge in their vernacular languages with AI assistance. Both avenues harness the collective intelligence of the community. The potential of the crowd is evident in the success of platforms such as Wikipedia \cite{surowiecki2005wisdom, kitturHarnessingWisdomCrowds2008}. However, the quality of crowd-sourced contributions is strongly influenced by the composition and motivation of participants; intentional engagement yields higher quality outcomes \cite{linWisdomCrowdsEffect2020}. Designing systems that build upon teachers’ existing Communities of Practice, where peer support and content sharing already happen, could strengthen both the correction and expansion of content. This would offer a scalable model that supplements expert review rather than replacing it.

\subsection{Limitations and Future Work}
Although our study is rigorous in terms of the methodology and participant demography, it has several limitations. First, the tool was deployed only during the partial academic year, from December 2024 to March 2025. 
Future deployments from the beginning of the school year may yield different insights into sustained use and adaptation. Second, despite the sizable sample (n = 1043), the study was confined to Karnataka, India. While many educational systems in the Global South share infrastructural and pedagogical traits, differences in language, policy, and local ecosystems may influence the extent to which the findings translate to other contexts. Expanding to other regions would offer valuable comparisons and insights into how context shapes teacher-AI collaboration. 
The scope of implementation was also limited to grades 5 through 10, which may not capture the distinct needs and practices of teachers at lower grade levels. Moreover, our analysis focused on teacher perceptions, log data, and AI-generated content, but did not examine the impact of the tool on student learning outcomes. While teachers reported reduced stress, improved efficiency, and enhanced pedagogical practices, it remains unclear how these changes translate into classroom learning. Future research should include longitudinal studies that examine student outcomes over time, and consider how sustained AI integration influences both teaching quality and learning trajectories. 

\section{Conclusion}
This study examined how teachers in low-resource classrooms integrate an AI-based lesson planning tool into their instructional practices. By deploying the tool with over 1,000 teachers, we gained valuable insights into how AI-generated LPs transform teaching workflows in challenging environments. Our findings reveal that the AI system significantly reduced administrative burden while simultaneously enhancing instructional quality. Teachers strategically leveraged the curriculum-aligned content as a foundation, enriching it with localized examples, interactive activities, and tailored assessments to meet their students' specific needs. The tool's mobile-optimized, low-tech design proved particularly effective within infrastructural constraints of low-resource schools, allowing teachers to create more engaging lessons with reduced preparation time.
The study also identified important limitations. Language support for vernacular content presented challenges, with Kannada LPs requiring substantial modifications compared to English materials. Additionally, while individual teachers benefited from the tool, its impact was constrained by broader systemic issues including staffing shortages and administrative pressures. The largely individual usage pattern, without sustained mentoring or collaborative frameworks, limited the tool's potential for deeper pedagogical transformation. These findings highlight that effective AI integration in education requires more than technological solutions. Success depends on embedding these tools within existing communities of practice where teachers can collectively adapt, reflect upon, and share knowledge about AI-assisted teaching. By positioning AI as a participatory resource aligned with teachers' values and expertise rather than a standalone solution, we can better support educational improvement in low-resource contexts. The tool has continued to evolve beyond the study period, with ongoing refinements shaped by the insights we gathered. It has also continued to scale to reach more teachers as we explore pathways to broaden its use in ways that remain responsive to their needs and contextual realities.

\section*{Acknowledgements}
\addcontentsline{toc}{section}{Acknowledgements}
We would like to thank the teachers who participated in the study for their time and thoughtful inputs. We extend our acknowledgments to Sikshana Foundation, especially Srinivas Mysore and Rijutha Raj, for their support in conducting the study. We thank Global Cornell and Cornell Global AI initiative for supporting this work. We also thank anonymous reviewers for their constructive and respectful feedback on our work.

\bibliographystyle{ACM-Reference-Format}
\bibliography{references}

\clearpage
\appendix

\section{Supplementary Figures}

\begin{figure}[ht]
    \centering
    \caption{Outcomes from the intervention}
    \scriptsize{Density distributions comparing pre-intervention (purple) and post-intervention (green) measurements based on teacher survey responses.}
    \includegraphics[width=\linewidth]{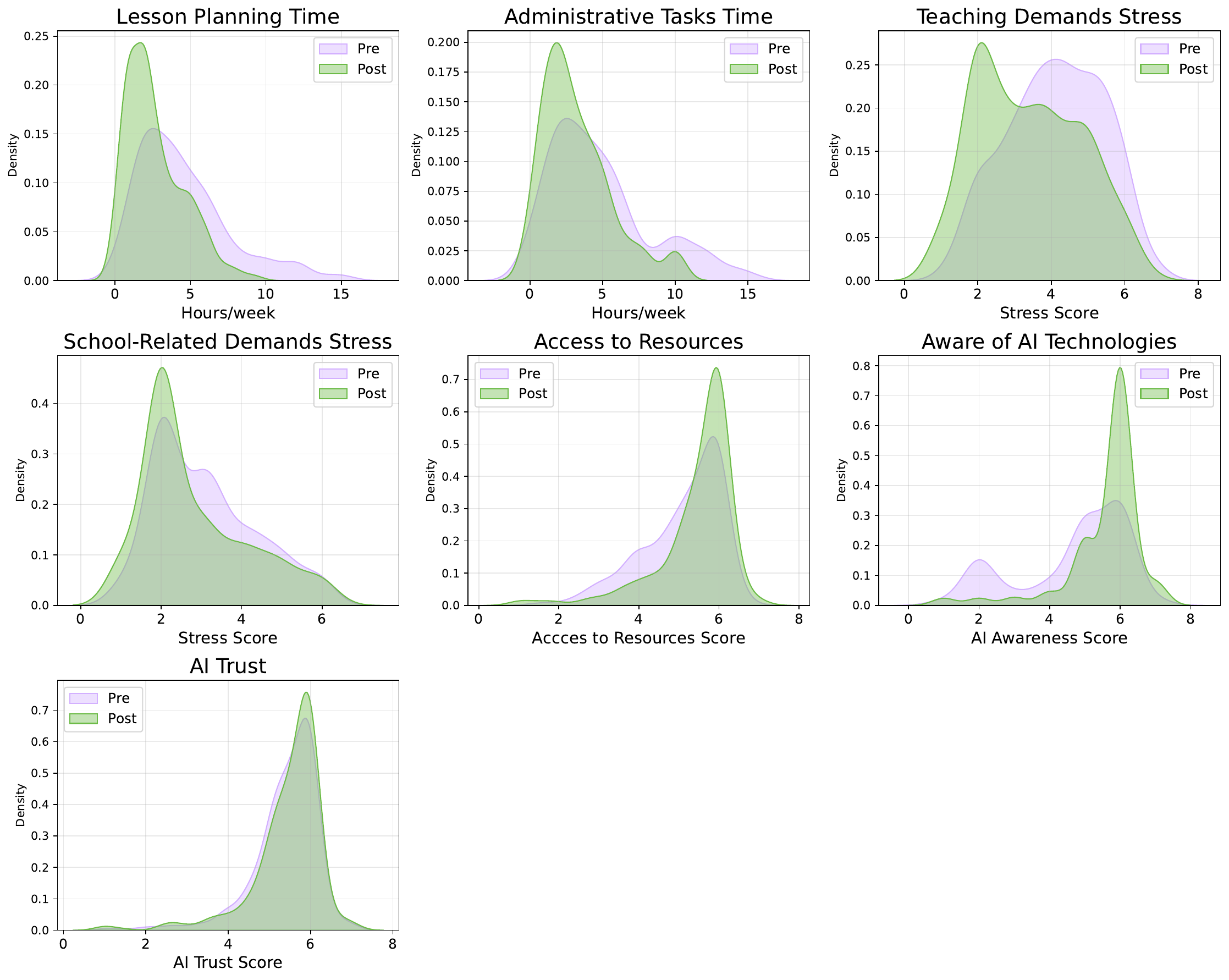}
    \label{fig:intervention-outcomes}
\end{figure}

\begin{figure}[ht]
    \centering
    \caption{Teacher Engagement, Lesson Planning Time, and Stress Scores}
    \scriptsize{From left to right: 
    A) Number of days teachers engaged with the tool versus the number of lesson plans they created; 
    B) Lesson planning time before and after the intervention, shown separately for less active and active users; 
    C) Teaching-related stress scores before and after the intervention, again shown separately for less active and active users.}
    \includegraphics[width=\linewidth]{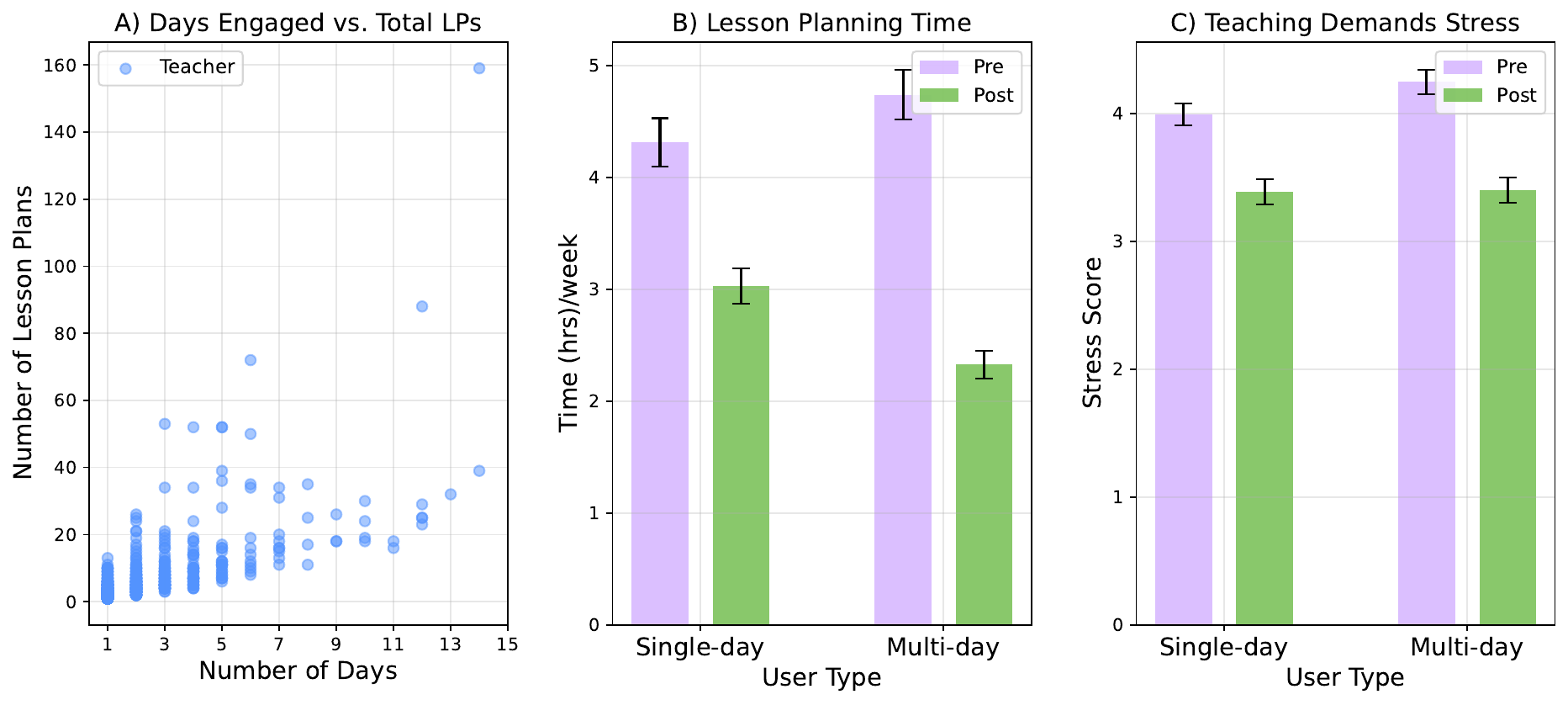}

    \label{fig:engagement-and-impact-based-on-activity}
\end{figure}

\begin{figure}[ht]
    \centering
    \begin{minipage}{0.49\columnwidth}
        \centering
        \caption{Resource Usage Comparison}
        \scriptsize{Usage of resources available in the lesson plans for supporting their teaching as reported by the teachers.}
        \includegraphics[width=\linewidth]{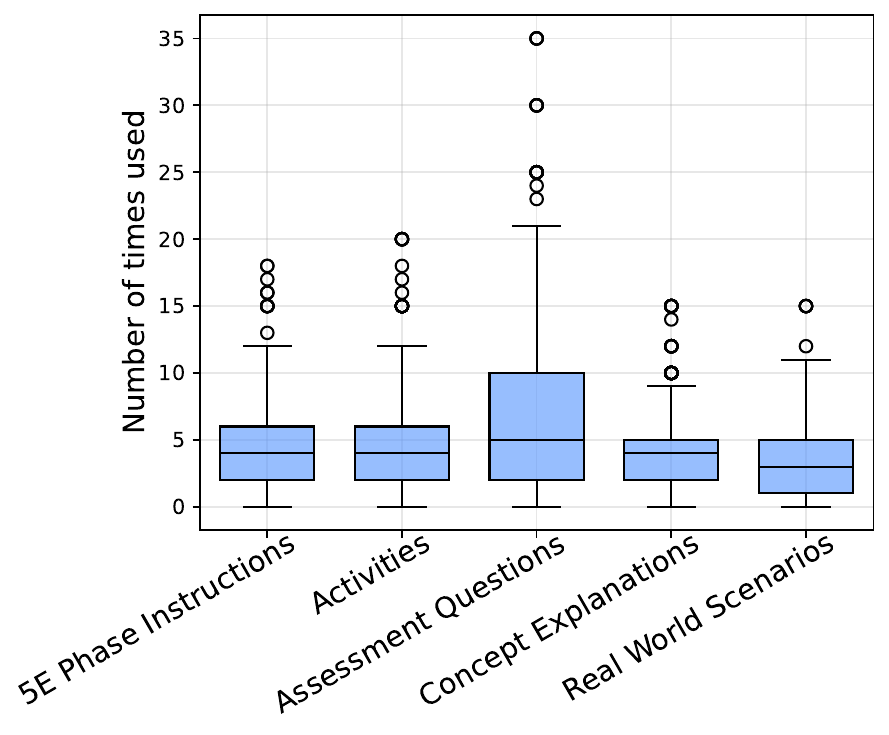}

        \label{fig:resource-usage}
    \end{minipage}%
    \hfill
    \begin{minipage}{0.49\columnwidth}
        \centering
        \caption{Extra Time Spent After LP Generation}
        \scriptsize{The extra time spent on documentation and teaching preparation by the teachers after the generation of the lesson plans in the tool.}
        \includegraphics[width=\linewidth]{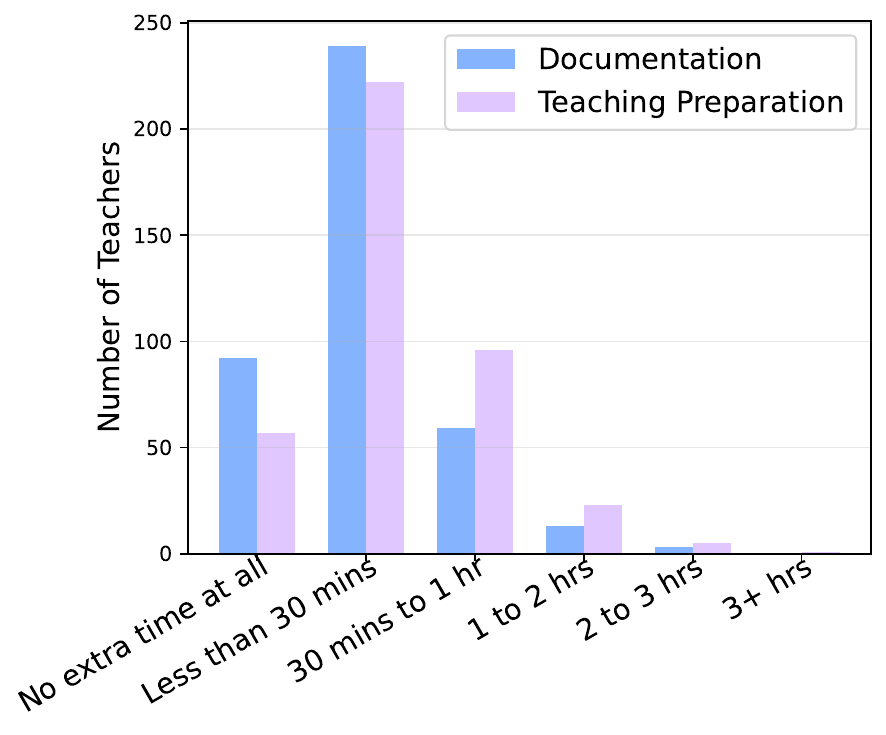}

        \label{fig:extra-time}
    \end{minipage}
\end{figure}

\begin{table}[t]
    \centering
    \caption{Example of AI-generated vs curated Learning Objectives from a Science Lesson }
    \begin{tabularx}{0.98\textwidth}{|>{\raggedright\arraybackslash}X|>{\raggedright\arraybackslash}X|}
\hline

\textbf{AI-generated Learning Objectives} & \textbf{Curator-edited Learning Objectives} \\ \hline
\textbf{   } \newline
\textbf{9.1 REFLECTION OF LIGHT} 
\begin{itemize}
\item \hlred{Understand} the \hlred{concept of reflection of light}.
\item \hlred{Explain} the \hlred{laws} of reflection of light.
\item Describe the \hlred{behavior of light on polished surfaces like mirrors}. 
\end{itemize}

&
\textbf{   } \newline
\textbf{9.1 REFLECTION OF LIGHT} 
\begin{itemize}
  \item \hlgreen{State and explain} the \hlgreen {laws of reflection for plane surfaces}.
  \item \hlgreen{Identify} the \hlgreen{properties} of reflection of light.
  \item Describe the \hlgreen{image formed by a plane mirror (virtual, erect, same size, laterally inverted)}.
  \item \hlgreen{Explain the relationship between the object distance and image distance for a plane mirror.}
  \item \hlgreen{Differentiate between concave and convex spherical mirrors based on their shape.}
  \item \hlgreen{Describe the general characteristics of images formed by concave and convex mirrors (real/virtual, erect/inverted, magnified/diminished).}
  \item \hlgreen{Discuss the uses of concave mirrors and convex mirrors.}
\end{itemize}

\\ \hline
\end{tabularx}

    \label{tab:curator_edit_comparison}
\end{table}

\clearpage

\section{Codebook From Qualitative Analysis of Interview Data}\label{codebook}
\begin{table*}[ht]
\centering
\caption{The complete codebook that resulted from our analysis of qualitative interviews, showing five themes (bold) and 28 codes, and the total count for each theme/code. The count for each theme is the sum of the counts of all codes within that theme.}
\begin{tabular}{|p{7.5cm}|c|}
\hline
\textbf{Theme/Code} & \textbf{Count} \\
\hline
\textbf{Quality of Lesson Plans} & \textbf{204} \\
\hline
Good Quality Lesson Plans & 45 \\
Flexibility to pick from multiple activities & 30 \\
Request for Visual Material & 29 \\
Catering to Different Student Levels & 29 \\
Easy to Create Assessments & 22 \\
Name Ownership & 17 \\
Knowledge Enriching & 14 \\
Inclusive Classrooms  & 9 \\
Innovative Ideas & 5 \\
Refreshes Subject Knowledge & 4 \\

\hline
\textbf{Implementation Factors} & \textbf{114} \\
\hline
Resource Constraints & 25 \\
Depends on Teachers & 25 \\
Classroom Constraints & 24 \\
Real Time Curriculum Adjustment & 21 \\
Time Constraints & 19 \\

\hline
\textbf{Impact on Teaching Practices} & \textbf{142} \\
\hline
Reduction in Workload & 37 \\
Stress Reduction & 28 \\
Fulfill Bureaucratic Needs & 21 \\
Increase in Creativity & 20 \\
Increased Student Engagement & 20 \\
Change in Teaching Process & 16 \\

\hline
\textbf{Cultural Relevance and Bias} & \textbf{91} \\
\hline
Tied to Karnataka Context & 48 \\
Tied to the Local Context & 28 \\
Lack of Global Context & 15 \\
\hline
\textbf{AI Knowledge \& Perception} & \textbf{128} \\
\hline
Trust in AI & 45 \\
Over-reliance on AI & 32 \\
Potential in Supporting Education & 31 \\
Perception of AI & 20 \\
\hline
\end{tabular}
\label{tab:theme_counts}
\end{table*}

\clearpage

\section{Codebooks from Chatlogs}\label{chatlog-codebook}
\begin{table}[ht]
\centering
\caption{Thematic Codebook from Lesson Plan Chatlogs (Counts reflect chat queries, not conversations).}
\begin{tabular}{|p{10cm}@{}|c|}
\hline
\textbf{Theme/Code} & \textbf{Count} \\
\hline
\textbf{Assessment Generation} \newline MCQs, fill-in-the-blanks, match-the-following, one/two-mark, and exam questions. & 160 \\
\hline
\textbf{Worksheet \& Puzzle Requests} \newline Worksheet creation, slap-the-board, agree/disagree, riddles, and puzzles. & 30 \\
\hline
\textbf{Multilingual/Language-Specific} \newline Requests in Kannada or for Kannada translations. & 25 \\
\hline
\textbf{Lesson Plan Support} \newline Uploading/finding lesson plans or resolving format issues. & 20 \\
\hline
\textbf{Grammar \& Language Help} \newline Topics like adjectives, prepositions, rhyming, sentence structure. & 20 \\
\hline
\textbf{Literature and Poetry} \newline Poems, poets, and poem comprehension exercises. & 15 \\
\hline
\textbf{Concept Clarification} \newline Definitions, explanations, differences, examples. & 15 \\
\hline
\textbf{Video and Visuals Requests} \newline Videos, graphs, images, and charts. & 12 \\
\hline
\end{tabular}
\end{table}

\begin{table}[ht]
\centering
\caption{Thematic Codebook from General Chatbot Queries (Counts reflect chat queries, not conversations).}
\begin{tabular}{|p{10cm}@{}|c|}
\hline
\textbf{Theme/Code} & \textbf{Count} \\
\hline
\textbf{Assessment Generation} \newline MCQs, one/two-mark, fill-in-the-blanks, match-the-following, exam questions. & 110 \\
\hline
\textbf{Multilingual Support} \newline Kannada-specific queries or translation requests. & 45 \\
\hline
\textbf{Lesson Plans \& Notes} \newline Requests for lesson plans, notes, summaries, chapter-wise materials. & 35 \\
\hline
\textbf{Concept Clarification} \newline Definitions, differences, examples, topic explanations. & 20 \\
\hline
\textbf{Grammar \& Language Help} \newline Grammar rules, punctuation, homophones, sentence types. & 18 \\
\hline
\textbf{Science Experiments \& Activities} \newline Demonstrations, classroom experiments, student activities. & 16 \\
\hline
\textbf{Teaching Aids \& Instruction} \newline Sample cards, classroom instruction, teaching aids. & 14 \\
\hline
\textbf{Poetry \& Literature} \newline Poetic devices, summaries, interpretations. & 12 \\
\hline
\textbf{Values, Events, \& Speeches} \newline Republic Day, Independence Day, moral/public speeches. & 10 \\
\hline
\end{tabular}
\end{table}

\clearpage

\section{Survey}\label{survey}

\noindent\textbf{Pre Survey Questionnaire}

\vspace{0.8em}
\noindent\textbf{C.1 Time Spent}

\textit{Instructions. Think about your teaching experience over the past two months:}

\noindent\hangindent=1.5em\hangafter=0 \textbf{Q1.1} How many hours per week do you spend creating lesson plans? \\
\noindent\hangindent=1.5em\hangafter=0 \textbf{Q1.2} How many hours per week do you spend on administrative tasks?

\vspace{0.8em}

\noindent\textbf{C.2 Stress}

\textit{Instructions. Please indicate how much you agree or disagree with the following statements:}

\noindent\hangindent=1.5em\hangafter=0 \textbf{Q2.1} I felt stressed for having too much teaching work to do. \\
\noindent\hangindent=1.5em\hangafter=0 \textbf{Q2.2} I felt stressed for not having enough time to complete my teaching work. \\
\noindent\hangindent=1.5em\hangafter=0 \textbf{Q2.3} I felt stressed for not being able to meet the diverse learning needs of my students. \\
\noindent\hangindent=1.5em\hangafter=0 \textbf{Q2.4} I felt stressed about not doing a good job with my teaching. \\

\textit{Response options:} 
Strongly disagree (1), Disagree (2), Somewhat disagree (3), Neither agree nor disagree (4), Somewhat agree (5), Agree (6), Strongly agree (7)

\vspace{0.8em}

\noindent\textbf{C.3 Resource Access}

\textit{Instructions. Please indicate how much you agree or disagree with the following statements:}

\noindent\hangindent=1.5em\hangafter=0 \textbf{Q3.1} I felt stressed for not having support from the administrators at my school. \\
\noindent\hangindent=1.5em\hangafter=0 \textbf{Q3.2} I felt stressed for not having support from colleagues at my school. \\
\noindent\hangindent=1.5em\hangafter=0 \textbf{Q3.3} I felt stressed for having to manage student behaviors. \\

\textit{Response options:} 
Strongly disagree (1), Disagree (2), Somewhat disagree (3), Neither agree nor disagree (4), Somewhat agree (5), Agree (6), Strongly agree (7)

\vspace{0.8em}

\noindent\textbf{C.4 Interactive Resources}

\textit{Instructions. Think about the teaching resources you've used in the past two months:}

\noindent\hangindent=1.5em\hangafter=0 \textbf{Q4.1} I have easy access to interactive classroom activities. \\
\noindent\hangindent=1.5em\hangafter=0 \textbf{Q4.2} It is easy to incorporate interactive activities while teaching. \\
\noindent\hangindent=1.5em\hangafter=0 \textbf{Q4.3} I have access to teaching plans contextualized to the local culture. \\

\textit{Response options:} 
Strongly disagree (1), Disagree (2), Somewhat disagree (3), Neither agree nor disagree (4), Somewhat agree (5), Agree (6), Strongly agree (7)

\vspace{0.8em}

\noindent\textbf{C.5 AI Usage \& Access}

\textit{Instructions. Please indicate how much you agree or disagree with the following statements:}

\noindent\hangindent=1.5em\hangafter=0 \textbf{Q5.1} I have easy access to digital teaching resources such as videos. \\
\noindent\hangindent=1.5em\hangafter=0 \textbf{Q5.2} I have easy access to assessment questions for students of different learning levels. \\
\noindent\hangindent=1.5em\hangafter=0 \textbf{Q5.3} I have access to teaching plans with real-world scenarios that connect textbook concepts to real life. \\
\noindent\hangindent=1.5em\hangafter=0 \textbf{Q5.4} I am aware of how AI technologies work. \\

\textit{Response options:} 
Strongly disagree (1), Disagree (2), Somewhat disagree (3), Neither agree nor disagree (4), Somewhat agree (5), Agree (6), Strongly agree (7) \\

\noindent\hangindent=1.5em\hangafter=0 \textbf{Q5.5} How often do you use the following applications? (Applications: ChatGPT, Meta AI)

\textit{Response options:} 
\textit{Never, Daily, Few times a week, Few times a month, Few times a year}

\vspace{0.8em}

\noindent\textbf{C.6 AI Trust Statements}

\textit{Instructions. Please indicate how much you agree or disagree with the following statements:}

\noindent\hangindent=1.5em\hangafter=0 \textbf{Q6.1} I trust using AI-based educational technology in my classroom. \\
\noindent\hangindent=1.5em\hangafter=0 \textbf{Q6.2} I believe I can successfully use AI-based technology in my classroom when available. \\
\noindent\hangindent=1.5em\hangafter=0 \textbf{Q6.3} I will use AI-based technology in my class when available. \\
\noindent\hangindent=1.5em\hangafter=0 \textbf{Q6.4} Using AI-based tools will require major changes to my teaching. \\
\noindent\hangindent=1.5em\hangafter=0 \textbf{Q6.5} As AI tools become common, I will trust them more. \\
\noindent\hangindent=1.5em\hangafter=0 \textbf{Q6.6} The more I understand how an AI tool makes decisions, the more I will trust it. \\
\noindent\hangindent=1.5em\hangafter=0 \textbf{Q6.7} The more data an AI tool has, the more I will trust its insights. \\
\noindent\hangindent=1.5em\hangafter=0 \textbf{Q6.8} I trust AI recommendations as much as I trust recommendations from a fellow teacher. \\
\noindent\hangindent=1.5em\hangafter=0 \textbf{Q6.9} I trust AI recommendations as much as I trust recommendations from an expert. \\

\textit{Response options:} 
Strongly disagree (1), Disagree (2), Somewhat disagree (3), Neither agree nor disagree (4), Somewhat agree (5), Agree (6), Strongly agree (7)

\vspace{1em}

\noindent\textbf{Post Survey Questionnaire}

\noindent\textit{Note: The Post Survey included all items from the Pre Survey Questionnaire, with the following additional section focused on tool usage.}

\vspace{0.8em}

\noindent\textbf{C.7 Utilization of \toolname}

\textit{Instructions. Think of your experience using \toolname and answer the following questions:}

\noindent\hangindent=1.5em\hangafter=0 \textbf{Q7.1} How many lesson plans from \toolname have you used for official documentation (e.g., submitting to supervisors)? \\
\noindent\hangindent=1.5em\hangafter=0 \textbf{Q7.2} How much extra time (on average) did you spend preparing \toolname lesson plans for documentation? \\
\noindent\hangindent=1.5em\hangafter=0 \textbf{Q7.3} Once you have created a lesson plan on \toolname, how much extra time (on average) did you spend on teaching preparation? \\

\textit{Response options:} 
No extra time at all, Less than 30 mins, 30 mins to 1 hr, 1 to 2 hrs, 2 to 3 hrs, 3+ hrs.
\\

\noindent\hangindent=1.5em\hangafter=0 \textbf{Q7.4} Overall how many of the following resources from \toolname did you implement in your classroom teaching? \\

\textit{Response options:} 5E Phase Instructions, Activities, Assessment Questions, Concept Explanations, Real-world Scenarios.\\

\clearpage

\section{Example Lesson Plan}\label{appendix:sample-lp}
\subsection{Lesson Plan format for Administrative Submission}

\begin{tabular}{|p{0.9cm}|p{1.3cm}|p{1.6cm}|p{2.7cm}|p{2.8cm}|p{2.5cm}|}
\hline
\textbf{Grade} & \textbf{Subject} & \textbf{LP Level} & \textbf{Chapter} & \textbf{Subtopics} & \textbf{Teacher Name} \\
\hline
10 & science\_2 & SUBTOPIC & Light - Reflection and Refraction & 9.1 REFLECTION OF LIGHT & [Anonymized] \\
\hline
\end{tabular}

\vspace{0.5em}

\noindent\textbf{Learning Outcomes:}
\begin{itemize}
    \item State and explain the laws of reflection for plane surfaces.
    \item Identify the properties of the image formed by a plane mirror (virtual, erect, same size, laterally inverted).
    \item Explain the relationship between the object distance and image distance for a plane mirror.
    \item Differentiate between concave and convex spherical mirrors based on their shape.
    \item Describe the general characteristics of images formed by concave and convex mirrors (real/virtual, erect/inverted, magnified/diminished).
    \item Discuss the uses of concave mirrors and convex mirrors.
\end{itemize}
{\small 
\begin{longtable}{|p{1.6cm}|p{5.5cm}|p{2.2cm}|p{1.8cm}|p{1.3cm}|}
\hline
\textbf{Phase} & \textbf{Classroom Process} & \textbf{TLM} & \textbf{CCE Tools and Techniques} & \textbf{Teacher Reflection} \\
\hline
\endfirsthead
\hline
\textbf{Phase} & \textbf{Classroom Process} & \textbf{TLM} & \textbf{CCE Tools and Techniques} & \textbf{Teacher Reflection} \\
\hline
\endhead

ENGAGE & Introduction to the concept of reflection using a captivating question about how mirrors work, followed by a discussion on the laws of reflection and the properties of images formed by different mirrors. Use a real-world scenario like signaling for help using a mirror or observing magnified images with a spoon. & Small mirrors, a spoon, a flashlight & Observation & \\
\hline
EXPLORE & Students will explore the effects of changing the angle of incidence on a plane mirror, the differences in image size and orientation when viewing through different parts of a spoon, and how distance affects the size and clarity of reflections in a plane mirror. & Small mirrors, a spoon, a flashlight, white wall or sheet & Observation & \\
\hline
EXPLAIN & Provide detailed explanations on the laws of reflection, properties of plane mirror images, the relationship between object and image distance, differences between concave and convex mirrors, and characteristics of images formed by spherical mirrors. & Diagrams showing the path of light reflecting off mirrors, examples of real-world applications & Discussion & \\
\hline
ELABORATE & Engage students in real-world scenarios and interactive activities like the Mirror Maze Challenge for smaller classrooms or Reflection Relay for larger classrooms, to apply the concepts learned in new situations. & Large sheet of paper, laser pointer, concave and convex mirrors, flashlight & Observation / Discussion & \\
\hline
EVALUATE & Use the provided question bank to assess students’ understanding of the topic through multiple-choice questions and assessment questions ranging from beginner to advanced levels. & Question bank with multiple-choice and assessment questions & Questionnaire & \\
\hline
\end{longtable}
}

\subsection{Detailed Lesson Plan For Teaching}

\begin{tabular}{|p{1.6cm}|p{5cm}|p{1.8cm}|p{4cm}|}
\hline
\textbf{Board} & KSEEB & \textbf{Medium} & English \\
\hline
\textbf{Class} & 10 & \textbf{Subject} & Science Sem2 \\
\hline
\textbf{Chapter} & 9. Light - Reflection and Refraction & \textbf{Sub-Topic} & 9.1 REFLECTION OF LIGHT \\
\hline
\end{tabular}

\vspace{1em}

\noindent\textbf{LESSON PLAN SUMMARY}

\noindent\textit{ENGAGE}

\noindent\textbf{Activity:} Introduction to the concept of reflection using a captivating question about how mirrors work, followed by a discussion on the laws of reflection and the properties of images formed by different mirrors. Use a real-world scenario like signaling for help using a mirror or observing magnified images with a spoon.

\noindent\textbf{Materials:} Small mirrors, a spoon, a flashlight

\noindent\textit{EXPLORE}

\noindent\textbf{Activity:} Students will explore the effects of changing the angle of incidence on a plane mirror, the differences in image size and orientation when viewing through different parts of a spoon, and how distance affects the size and clarity of reflections in a plane mirror.

\noindent\textbf{Materials:} Small mirrors, a spoon, a flashlight, white wall or sheet

\noindent\textit{EXPLAIN}

\noindent\textbf{Activity:} Provide detailed explanations on the laws of reflection, properties of plane mirror images, the relationship between object and image distance, differences between concave and convex mirrors, and characteristics of images formed by spherical mirrors.

\noindent\textbf{Materials:} Diagrams showing the path of light reflecting off mirrors, examples of real-world applications

\noindent\textit{ELABORATE}

\noindent\textbf{Activity:} Engage students in real-world scenarios and interactive activities like the Mirror Maze Challenge for smaller classrooms or Reflection Relay for larger classrooms, to apply the concepts learned in new situations.

\noindent\textbf{Materials:} Large sheet of paper, laser pointer, concave and convex mirrors, flashlight

\noindent\textit{EVALUATE}

\noindent\textbf{Activity:} Use the provided question bank to assess students' understanding of the topic through multiple-choice questions and assessment questions ranging from beginner to advanced levels.

\noindent\textbf{Materials:} Question bank with multiple-choice and assessment questions

\vspace{1em}

\noindent\textbf{LESSON PLAN DETAILS}

\noindent\textbf{Learning Outcomes}

\begin{enumerate}
    \item State and explain the laws of reflection for plane surfaces.
    \item Identify the properties of the image formed by a plane mirror (virtual, erect, same size, laterally inverted).
    \item Explain the relationship between the object distance and image distance for a plane mirror.
    \item Differentiate between concave and convex spherical mirrors based on their shape.
    \item Describe the general characteristics of images formed by concave and convex mirrors (real/virtual, erect/inverted, magnified/diminished).
    \item Discuss the uses of concave mirrors and convex mirrors.
\end{enumerate}

\noindent\textbf{ENGAGE}

\noindent\textbf{Introduction:} Have you ever wondered how mirrors work or why your reflection looks back at you?

\noindent\textbf{Key Concepts:} Laws of reflection, properties of plane mirror images, object and image distance relationship, differences between concave and convex mirrors, characteristics of images formed by spherical mirrors.

\noindent\textbf{Real-world Scenario:} Imagine trying to signal for help using a mirror's reflection or using a spoon to see a magnified version of your face.

\noindent\textbf{Vocabulary:} Reflection, plane mirror, virtual image, erect, laterally inverted, concave mirror, convex mirror, real image, magnified, diminished.

\noindent\textbf{Classroom Process (Facilitating Activity):}

\begin{enumerate}
    \item \textbf{Preparation Steps:} Teachers should gather a few small mirrors, a spoon, a flashlight, and a plain white wall or sheet. They should also prepare a simple diagram showing the path of light reflecting off a mirror.
    \item \textbf{Required Materials:} Small mirrors, a spoon (one side convex, the other concave), a flashlight, and a white wall or sheet.
    \item \textbf{Obtaining Materials:} These materials can be easily found in most homes or purchased at a local store. Mirrors and spoons are common household items, while a flashlight can be borrowed if not readily available. The white wall or sheet serves as a projection surface, readily available in any classroom or home.
\end{enumerate}

\vspace{0.5em}

\noindent\textbf{EXPLORE}

\begin{enumerate}
    \item How does changing the angle of incidence on a plane mirror affect the angle of reflection and the appearance of the image?
    \item What happens to the size and orientation of your reflection when you view yourself in different parts of a spoon? Explore both the concave and convex sides.
    \item Investigate how the distance between you and a plane mirror affects the size and clarity of your reflection. Does moving closer or further away change how you perceive your image?
\end{enumerate}

\vspace{0.5em}

\noindent\textbf{EXPLAIN}

\textbf{Laws of Reflection:}

\begin{itemize}
    \item \textbf{Definition:} Fundamental principles governing the behavior of light when it reflects off surfaces.
    \item \textbf{Characteristics:} States that the angle of incidence equals the angle of reflection, and both angles lie on the same plane as the reflecting surface.  
    \item \textbf{Key Features and Facts:} This law applies to all reflective surfaces, including plane, concave, and convex mirrors, and is crucial for understanding image formation.
    
\end{itemize}

\textbf{Properties of Plane Mirror Images:}

\begin{itemize}
\item \textbf{Definition:} Qualities and attributes of images formed by plane mirrors.
\item \textbf{Characteristics:} Images are virtual (cannot be projected on a screen), erect (upright), of the same size as the object, and laterally inverted (left-right reversed).
\item \textbf{Key Features and Facts:} Despite the lateral inversion, plane mirrors are commonly used in daily life due to their ability to produce undistorted images.
\end{itemize}

\textbf{Object and Image Distance Relationship in Plane Mirrors:}

\begin{itemize}
\item \textbf{Definition:} The correlation between the distances of the object and the image from the mirror.
\item \textbf{Characteristics:} The distance of the image from the mirror is equal to the distance of the object from the mirror, but in the opposite direction.
\item \textbf{Key Features and Facts:} This relationship helps in locating the image formed by plane mirrors and is essential for understanding mirror equations.
\end{itemize}

\textbf{Differences Between Concave and Convex Mirrors:}

\begin{itemize}
\item \textbf{Definition:} Comparison based on the curvature of concave and convex mirrors.
\item \textbf{Characteristics:} Concave mirrors curve inward and can produce real or virtual images, while convex mirrors curve outward and always produce virtual images.
\item \textbf{Key Features and Facts:} Concave mirrors can magnify images and are used in telescopes, while convex mirrors provide a wider field of view and are used in vehicle rear-view mirrors.
\end{itemize}

\textbf{Characteristics of Images Formed by Spherical Mirrors:}

\begin{itemize}
\item \textbf{Definition:} Description of the nature of images produced by concave and convex mirrors.
\item \textbf{Characteristics:} Images can be real or virtual, erect or inverted, and magnified or diminished, depending on the mirror type and object position.
\item \textbf{Key Features and Facts:} Concave mirrors can produce magnified real images, useful in applications like shaving mirrors, while convex mirrors always produce diminished images, useful for safety in vehicles.
\end{itemize}

\textbf{Uses of Concave and Convex Mirrors:}

\begin{itemize}
\item \textbf{Definition:} Practical applications of concave and convex mirrors in various fields.
\item \textbf{Characteristics:} Concave mirrors are used in devices requiring magnified images, while convex mirrors are used for wide-angle viewing.
\item \textbf{Key Features and Facts:} Concave mirrors are found in telescopes, headlamps, and solar furnaces, whereas convex mirrors are used in vehicle rear-view mirrors and security purposes.
\end{itemize}

\textbf{Classroom Process (Facilitating Activity):}

\begin{enumerate}
    \item \textbf{Detailed Steps for Implementation:} Students will use small mirrors, a spoon, and a flashlight to explore the laws of reflection, comparing the images formed by plane, concave, and convex mirrors. They will observe the differences in image size, orientation, and type (real or virtual) by changing the object's distance from the mirror and using different parts of the spoon.
    \item \textbf{Recap at the End of Activity:} Summarize the activity by discussing the observed properties of images formed by different mirrors, emphasizing the laws of reflection, the relationship between object and image distance, and the practical uses of concave and convex mirrors. Highlight how these principles apply in real-world scenarios, like signaling for help or using mirrors in vehicles for safety.
\end{enumerate}

\vspace{0.5em}
\noindent\textbf{ELABORATE}

\textbf{Real-World Scenarios}

\begin{enumerate}
    \item \textbf{Scenario: Reflecting on History in Karnataka} \\
    \textit{Question:} How might the ancient architects of Karnataka have used the principles of reflection in designing structures like the reflective water pools in front of temples? \\
    \textit{Description:} These pools were strategically placed to reflect the majestic temple structures, utilizing the laws of reflection to enhance the beauty and grandeur of the temples. The water surface acts like a plane mirror, creating a virtual, erect, and laterally inverted image of the temple, demonstrating the properties of plane mirror images in a serene, natural setting.

    \item \textbf{Scenario: Navigating Through Karnataka's Roads} \\
    \textit{Question:} Why do drivers in the bustling streets of Bengaluru or the winding roads of the Western Ghats prefer convex mirrors for their vehicles? \\
    \textit{Description:} Convex mirrors provide a wider field of view, allowing drivers to see more of their surroundings, which is crucial for navigating through busy or curved roads. This practical use of convex mirrors showcases their ability to produce diminished, virtual images, ensuring safety by giving drivers a comprehensive view of traffic conditions.

    \item \textbf{Scenario: The Coastal Mirrors} \\
    \textit{Question:} How can fishermen on Karnataka's coast use concave mirrors to concentrate sunlight for signaling during the day? \\
    \textit{Description:} Fishermen can use handheld concave mirrors to focus sunlight into a small, intense beam directed towards a target for signaling. This application leverages the concave mirror's ability to converge light rays to a focal point, demonstrating its real-world utility in emergency signaling and the principles of light reflection.
\end{enumerate}

\textbf{Interactive Activities}

\begin{enumerate}
    \item \textbf{Collaborative Activity for Classrooms with Fewer Than 30 Students: Mirror Maze Challenge} \\
    - Students will work in small teams to create a maze on a large sheet of paper. They will then use small plane mirrors to direct a laser pointer's beam through the maze from start to finish, reflecting off the mirrors. This activity encourages students to apply the laws of reflection to navigate the maze successfully, fostering teamwork and problem-solving skills.

    \item \textbf{Group Activity for Classrooms with More Than 30 Students: Reflection Relay} \\
    - Divide the class into larger teams, and set up a relay race where each team must use a concave and a convex mirror to direct sunlight or a flashlight beam across a series of checkpoints. Each checkpoint will have a specific task, such as "use the convex mirror to broaden the light beam" or "use the concave mirror to focus the beam on a target." The first team to complete all tasks wins. This activity promotes understanding of the unique properties of concave and convex mirrors through a fun, competitive challenge.
\end{enumerate}

These activities are designed to deepen students' understanding of the reflection of light and the properties of different types of mirrors, extending their learning beyond the theoretical concepts into engaging, hands-on experiences that highlight the practical applications of these principles in everyday life and the natural world.

\vspace{0.5em}

\noindent\textbf{EVALUATE}

\noindent\textit{Beginner}

\textbf{MCQs}

\begin{enumerate}
    \item What does the first law of reflection state? \\
    A) The angle of incidence is greater than the angle of reflection. \\
    B) The angle of incidence is less than the angle of reflection. \\
    C) The angle of incidence is equal to the angle of reflection. \\
    D) The angle of incidence has no relation to the angle of reflection.
    
    \item How is the image formed by a plane mirror? \\
    A) Real and inverted \\
    B) Virtual and erect \\
    C) Real and erect \\
    D) Virtual and inverted
\end{enumerate}

\textbf{Assessments}

\begin{enumerate}
    \item Explain the laws of reflection for plane surfaces in your own words.
    \item Describe the properties of the image formed by a plane mirror.
\end{enumerate}

\noindent\textit{Intermediate}

\textbf{MCQs}

\begin{enumerate}
    \item If an object is placed 10 cm in front of a plane mirror, how far is the image from the mirror? \\
    A) 5 cm \\
    B) 10 cm \\
    C) 20 cm \\
    D) 15 cm

    \item Which mirror can form a real image? \\
    A) Plane mirror \\
    B) Convex mirror \\
    C) Concave mirror \\
    D) Both convex and concave mirrors
\end{enumerate}

\textbf{Assessments}

\begin{enumerate}
    \item Discuss how the distance of the object from a plane mirror affects the image distance and size.
    \item Differentiate between concave and convex spherical mirrors based on their shape and the types of images they form.
\end{enumerate}

\noindent\textit{Advanced}

\textbf{MCQs}

\begin{enumerate}
    \item A concave mirror can form a virtual, erect, and magnified image when the object is placed: \\
    A) At the focus \\
    B) Between the focus and the mirror \\
    C) Beyond the center of curvature \\
    D) At the center of curvature

    \item Which of the following is not a use of convex mirrors? \\
    A) In car side mirrors for a wider field of view \\
    B) In telescopes to gather more light \\
    C) In security mirrors in stores \\
    D) In shaving mirrors to see a larger image of the face
\end{enumerate}

\textbf{Assessments}

\begin{enumerate}
    \item Given a scenario where a concave mirror is used as a dentist's mirror, explain how the mirror helps the dentist see enlarged images of the patient's teeth.
    \item Discuss the uses of convex mirrors in daily life and explain why they are suited for these applications.
\end{enumerate}
\begin{enumerate}
    \item Discuss the uses of convex mirrors in daily life and explain why they are suited for these applications.
    \end{enumerate}
\end{document}